\documentclass
[10pt,aps,prb,superscriptaddress,reprint,showpacs,amssymb,amsmath,floatfix,citeautoscript,longbibliography]{revtex4-2}

\usepackage{graphicx} %include figure files
\usepackage{dcolumn} %align table columns on decimal point
\usepackage{bm} %bold math

\usepackage{ulem}
\usepackage[]{color}

\begin{document}
%\preprint{APS/123-QED}

%Title of paper
\title{Quantitative investigation of the 4$f$ occupation in the quasikagome Kondo lattice CeRh$_{1-x}$Pd$_x$Sn}

\author{Martin~Sundermann}
    \affiliation{Max Planck Institute for Chemical Physics of Solids, N{\"o}thnitzer Str. 40, 01187 Dresden, Germany}
    \affiliation{Deutsches Elektronen-Synchrotron DESY, 22607 Hamburg, Germany}
\author{Andrea~Marino}
    \affiliation{Max Planck Institute for Chemical Physics of Solids, N{\"o}thnitzer Str. 40, 01187 Dresden, Germany}
\author{Andrei~Gloskovskii}
    \affiliation{Deutsches Elektronen-Synchrotron DESY, 22607 Hamburg, Germany}
\author{Chongli~Yang}
    \altaffiliation{present address: Beijing Academy of Quantum Information Sciences, Beijing 100193, China}    
    \affiliation{Graduate School of Advanced Science and Engineering, Hiroshima University, Higashi-Hiroshima, 739-8530, Japan}
\author{Yasuyuki~Shimura}
    \affiliation{Graduate School of Advanced Science and Engineering, Hiroshima University, Higashi-Hiroshima, 739-8530, Japan}
\author{Toshiro~Takabatake}
    \affiliation{Graduate School of Advanced Science and Engineering, Hiroshima University, Higashi-Hiroshima, 739-8530, Japan}
\author{Andrea~Severing}
    \affiliation{Institute of Physics II, University of Cologne, Z\"{u}lpicher Str. 77, D-50937 Cologne, Germany}

\date{\today}

\begin{abstract}
CeRhSn with the Ce atoms forming a quasikagome lattice in the hexagonal plane has recently been discussed in the context of quantum criticality driven by magnetic frustration. Furthermore, it has been reported that the successive substitution of Rh by Pd leads to magnetic order.  Here we have investigated the change of the 4$f$ occupation in the substitution series CeRh$_{1-x}$Pd$_x$Sn  for for $x$\,=\,0, 0.1, 0.3, 0.5, 0.75 by means of photoelectron spectroscopy with hard x-rays (HAXPES). The quantitative analysis of the core level spectra with a combined full multiplet and configuration interaction analysis shows a smooth decrease of the 4$f^0$ contribution with rising $x$ due to an increase of the effective 4$f$ binding energy $\varepsilon_{4f}$ and the reduction of the effective hybridization $V_\text{eff}$. We further compare valence band data with the calculated partial density of states and find that the Pd\,4$d$ states are about 1\,eV further away from the Ce\,4$f$ states at the Fermi energy than the Rh\,4$d$ states. In fact, the effective binding energy $\varepsilon_{4f}$ of the 4$f$ states in the configuration interaction analysis of the core level spectra decreases by the same amount. 

\end{abstract}


\maketitle

\section{Introduction}
The physics of heavy fermion or Kondo lattice systems is driven by the hybridization of localized 4$f$ and conduction electrons ($cf$-hybridization)\,\cite{Coleman2007,Coleman2015}. The $cf$-hybridization increases with the exchange interaction $\cal{J}$, thus leading to a competition of the RKKY (Ruderman-Kittel-Kasuya-Yosida) interaction that favors magnetic order and the Kondo interaction that leads to a non-magnetic ground state. Kondo screening is the dominant interaction for strong $cf$-hybridization and with increasing $\cal{J}$, eventually, magnetic order is suppressed to zero Kelvin\,\cite{Doniach1977} and a quantum critical point occurs\,\cite{Hilbert2007,Wirth2016}. With further increasing $cf$-hybridization the $f$ electrons are partially delocalized so that the occupation of the 4$f$ shell is no longer integer. A non-integer 4$f$ occupation usually goes along with a large Kondo temperature $T_K$, the latter determining the energy scale of the Kondo interaction.

\begin{figure}[t]
 \centering
		\includegraphics[width=0.8\columnwidth]{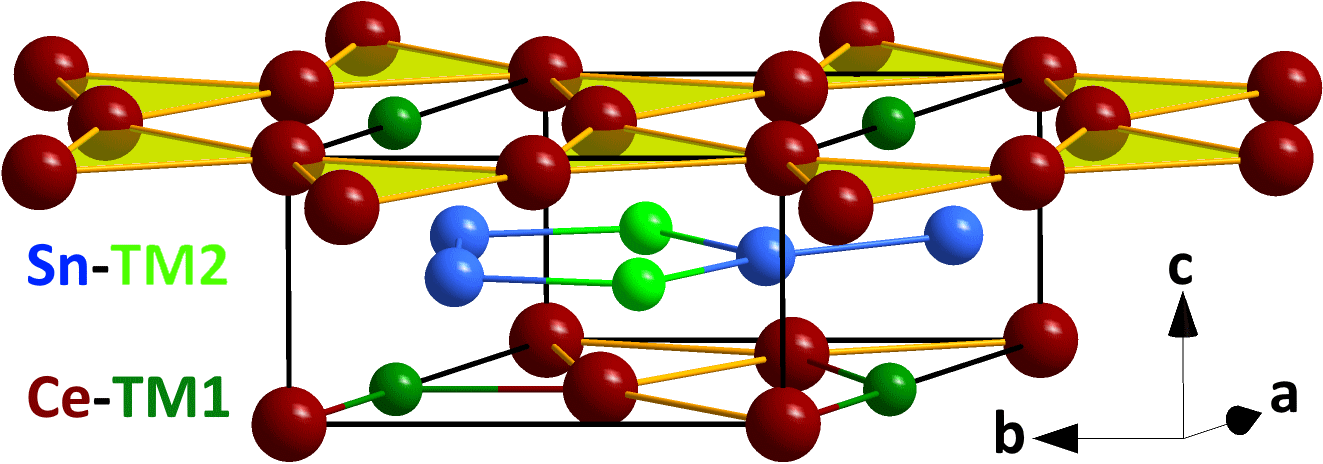} 
  \caption{(Color online) ZrNiAl-type structure of CeRhSn where the Ce atoms form a quasikagome lattice with alternating planes of Ce-Rh (TM1) and Sn-Rh (TM2). Structure data from Ref.\,\onlinecite{Kim2003}.} 
 \label{structure}
\end{figure}

More recently geometrical frustration has been suggested as another tuning parameter for approaching quantum criticality in strongly correlated electron materials\,\cite{Grigera2004}. This implies an additional axis for frustration in the phase diagram of intermetallic heavy fermion compounds\,\cite{Senthil2004,Si2006,Vojta2010,Coleman2010,Si2014}. Kagome lattices are prone to frustration\,\cite{Lacroix2010}, and indeed, in the intermetallic Kondo lattice compound CePdAl, which forms in the ZrNiAl-type structure (see Fig.\,\ref{structure}) where the Ce atoms form a quasikagome lattice in the hexagonal plane, effects of frustration have been observed\,\cite{Donni1996,Oyamada2008,Fritsch2014,Fritsch2017,Lucas2017}. CePdAl has a small Kondo temperature $T_K$ of 6\,K and orders antiferromagnetically below 2.7\,K with only two out of three Ce ions participating in this order. In addition short-range magnetic correlations have been observed\,\cite{Huesges2017,Zhao2019}. Both effects are interpreted in terms of frustration. CeRhSn crystallizes in the same hexagonal structure\,\cite{Pottgen2015} but, in contrast to CePdAl, it is a large $T_K$ system\,\cite{Nohara1993} ($T_K$\,=\,200\,K) and remains paramagnetic down to at least 50\,mK\,\cite{Kim2003,Schenck2004} with indications for the proximity to a magnetic quantum critical point\,\cite{Tokiwae2015}. The application of uniaxial stress or a magnetic field in the hexagonal plane seemingly pushes CeRhSn from a quantum critical state into a long-range ordered state\,\cite{Yang2017,Kuchler2017}. This is a great surprise because in cerium compounds pressure tends to suppress magnetic order (see e.g. in Ref.s\,\cite{Grosche2001} and \cite{Grube2018} and references therein). The formation of the magnetic ground state upon application of uniaxial pressure in the $ab$-plane has, therefore, been interpreted in terms of a stress induced reduction of geometrical frustration in the hexagonal plane.
 
%%%%%%%%%%%%%%%%%%%%%%%%%

%%%%%%%%%%%%%%%%%%%%%%%%%
\begin{figure*}[t]
 \centering
		\includegraphics[width=1.8\columnwidth]{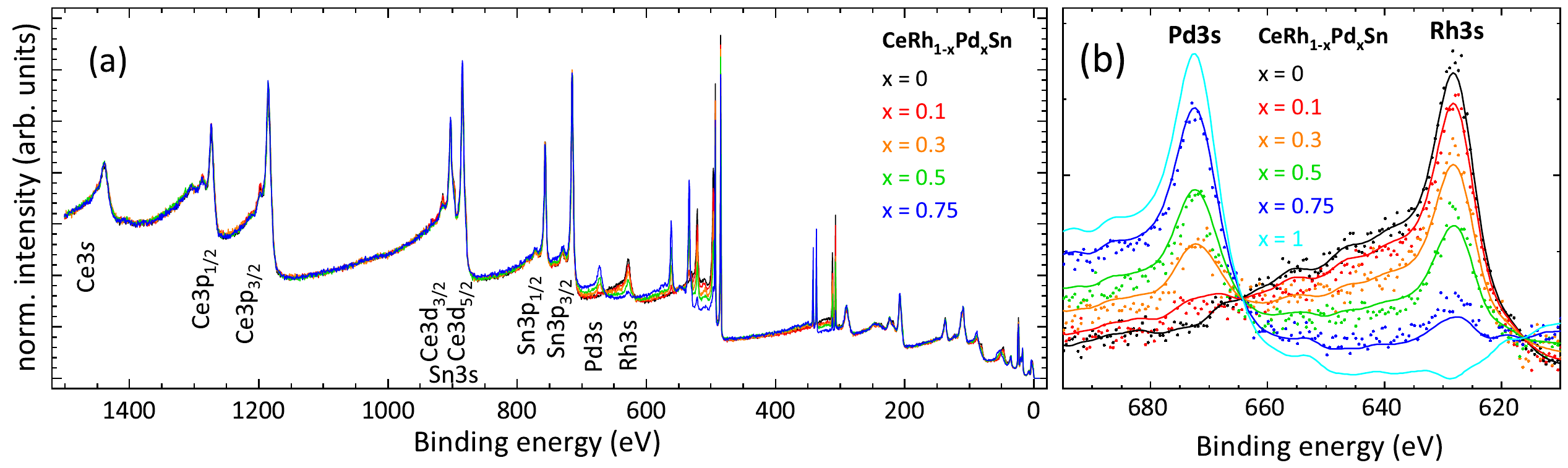} 
  \caption{(Color online) (a):\,Photo-electron emission scans of CeRh$_{1-x}$Pd$_x$Sn with $x$\,=\,0, 0.1, 0.3, 0.5, 0.75 over a wide energy range. (b) zoom into emission lines of Pd\,3$s$ and Rh\,3$s$, data (dots) and calculated intensities (lines, see text for details).}  
 \label{long}
\end{figure*}
%%%%%%%%%%%%%%%

Here we focus on the substitution series CeRh$_{1-x}$Pd$_x$Sn. Up to $x$\,$\approx$\,0.8, all compositions have the same hexagonal structure as CeRhSn. Here Pd occupies both Rh sites equally. For higher Pd concentrations the orthorhombic structure of CePdSn (TiNiSn-type) is adopted\,\cite{Niehaus2015}. Niehaus\,\textit{et al.}\,\cite{Niehaus2015} characterized the substitution series in great detail and found from the expanding volume, the magnetic susceptibility, and L-III edge absorption measurements that Pd stabilizes the trivalent state. Yang \textit{et al}. reported that the specific heat divided by temperature $C$/$T$ and the ac susceptibility develop peaks for $x$\,$\ge$\,0.1 which suggests the formation of a magnetically ordered ground state\,\cite{Yang2017}. The ordering temperature rises up to 2.7\,K for $x$\,=\,0.75. The static susceptibility and isothermal magnetization curves show an increase of the magnetic moment as the Pd content increases, while the Curie-Weiss temperature decreases, thus implying a decrease of the Kondo temperature $T_K$. Hence, the substitution with Pd drives CeRh$_{1-x}$Pd$_x$Sn  away from the quantum critical point in CeRhSn.  The question of the impact of frustration upon Pd substitution remains. It requires a detailed analysis of the putative magnetically ordered states with $\mu$SR and/or neutron diffraction\,\cite{Adroja} and the quantification of the of the $cf$-hybridization as a function of the Pd substitution $x$. The present manuscript addresses the Kondo interaction. 

For this purpose we present hard x-ray photoelectron spectroscopy (HAXPES) measurements of the Ce\,3$d$ and Ce\,3$p$ core level, and of the valence bands of CeRh$_{1-x}$Pd$_x$Sn with $x$\,=\,0, 0.1, 0.3, 0.5, and 0.75 with a quantitative analysis. Weakly hybridized 4$f$ states in Ce compounds are trivalent and have a 4$f$ occupation of $n_f$\,=\,1. Strong hybridization, on the other hand, leads to an intermixing of 4$f$ configurations so that the valence is no longer integer. The 4$f$ ground state is now a mixed state of the form $|\text{GS} \rangle$\,=\,$\alpha |f^0\rangle$\,+\,$\beta |f^1 \underline{\text{L}} \rangle$\,+\,$\gamma |f^2 \underline{\underline{\text{L}}} \rangle$, with $\underline{\text{L}}$ and $\underline{\underline{\text{L}}}$ standing for the number of ligand holes. Here the amount of $f^0$ quantifies the degree of the delocalization. In core level spectroscopy, the mixed ground state of the initial state is split up into three spectral weights 
$|\underline{\text{c}} f^2 \underline{\underline{\text{L}}} \rangle$, $|\underline{\text{c}} f^1 \underline{\text{L}} \rangle$, and $|\underline{\text{c}} f^0 \rangle$  in the final state due to the presence of the core hole\,\cite{Gunnarson2001}, and the corresponding spectral weights give information about $\alpha^2$, $\beta^2$, and $\gamma^2$. The relation between these spectral weights and the respective amounts of the different configurations in the wave function is not a simple proportionality due to hybridization effects in the final state, so that configuration interaction calculation (CI) is needed to relate the measured intensities to the $f^i$ contributions, $i$\,=\,0,1, and 2 in the initial state.

%>>>>>>>>>>>>>>>>>>>>>>>>>>>>>>>>>>>>>>>>>>>>>>>>>>>>>>>>>>>>

%%%%%%%%%%%%%%%%%%%%%%%%%
\begin{figure*}[]
 \centering
		\includegraphics[width=1.8\columnwidth]{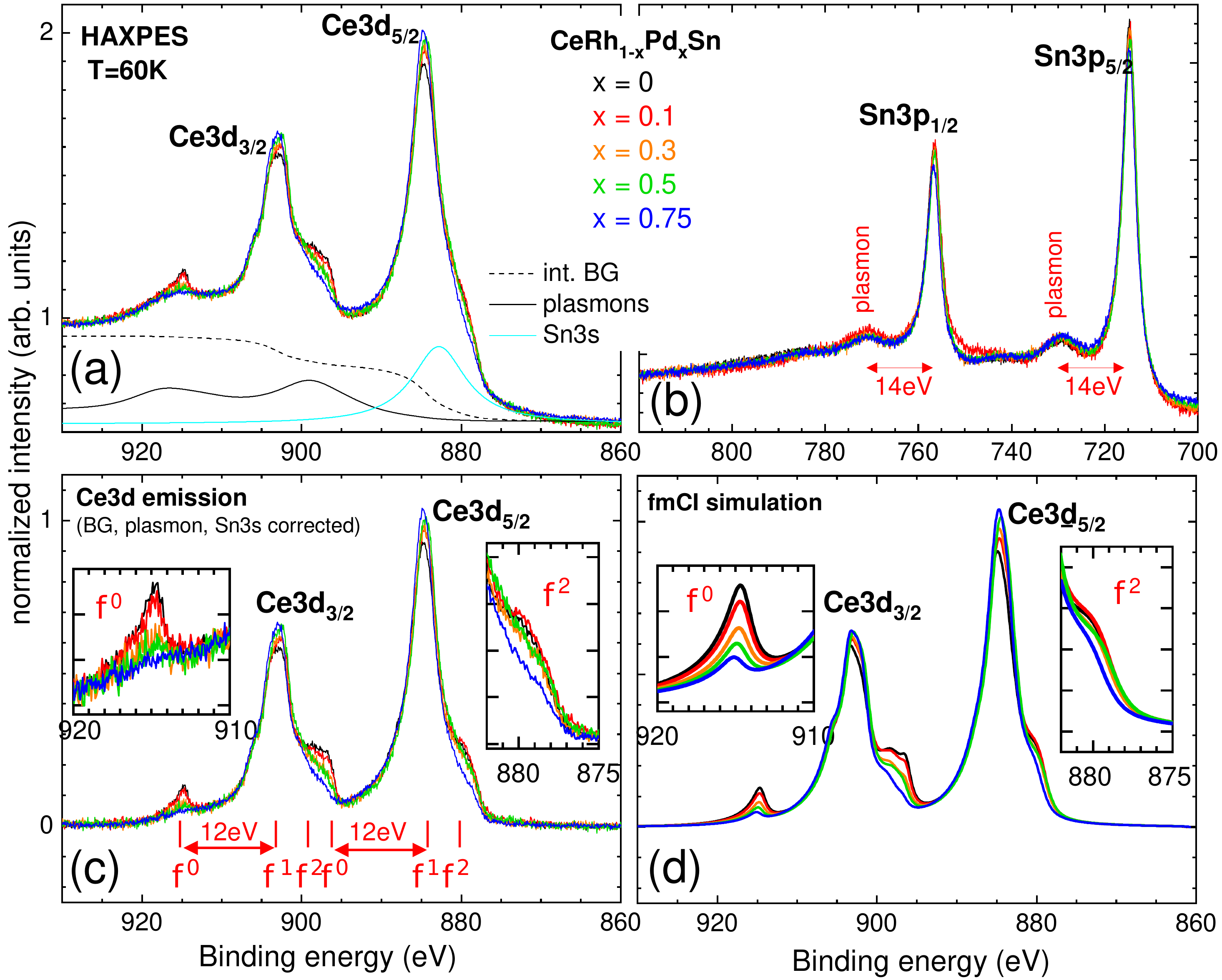} 
  \caption{(Color online) (a) Ce\,3$d$ core-level HAXPES data of CeRh$_{1-x}$Pd$_x$Sn with $x$\,=\,0, 0.1, 0.3, 0.5, 0.75. at $T$\,=\,60\,K after normalization to the integrated intensity before subtracting an integrated background (BG) (dashed black line), plasmons (solid black line), and the Sn\,3$s$ emission intensity (cyan line). BG and plasmons are specimen of the 30\%Pd analysis. (b) Sn\,3$p$ core level of all CeRh$_{1-x}$Pd$_x$Sn samples with plasmon intensities 14\,eV above the main emission lines (see red arrows). (c) Ce\,3$d$  emission lines after subtraction of BG, plasmons, and Sn\,3$s$. The red ruler indexes the center of the respective multiplet structures of the contributing 4$f$ configurations, $f^0$, $f^1$, and $f^2$. Insets: enlarged regions of $f^0$ and $f^2$ emission. (d) Results of the fmCI simulations for the Ce\,3$d$ emission using the CI parameters as listed in Table I.} 
 \label{Ce3d}
\end{figure*}
%%%%%%%%%%%%%%%

\section{Experiment and Simulation}

For the HAXPES experiments of  CeRh$_{1-x}$Pd$_x$Sn, single crystalline samples were used for $x$\,=\,0 and 0.1, while polycrystalline samples were used for higher $x$. However, it turned out that the polycrstalline samples consisted of long grains (0.5-1.0 mm) along the hexagonal $c$ axis. For the HAXPES measurements, clean surfaces perpendicular to the $c$ axis were prepared by breaking the sample bar in a high vacuum. The single crystals for $x$\,=\,0 and 0.1 were also cut perpendicular to the $c$ axis so that we do not expect that the single or poly-crystalline nature of the samples has an impact on the results. The methods of preparation and characterizations of samples are described in Ref.\,\onlinecite{Yang2017}. The electron-probe microanalysis showed that the samples with $x$ up to 0.5 are homogeneous but the actual composition for the sample with the initial composition of 0.8 is approximately 0.75 due to the formation of some trivalent Ce impurity phases (alloys of CeRh$_2$Sn$_2$ and CePd$_2$Sn$_2$)\,\cite{Beyermann1991} that amount in total to a maximum of 5\%\,\cite{Yang2017}. The CeIrSn sample that was used for the valence band measurements is the same as in Ref.\,\cite{Shimura2021}.

Photoelectron spectroscopy with soft x-rays (PES) has been shown to be a very valuable technique for the investigation of the electronic states of rare-earth compounds \cite{Allen1986,Huefner1992,Tjeng1993}, but suffers from surface effects. PES with hard x-rays (HAXPES), on the other hand, provides the bulk sensitivity of about 80\AA that is necessary to image the bulk electronic structure in these systems\,\cite{Braicovich1997,Dallera2005,Suga_Sm_2007,TanumaIMFP_2011}, especially in correlated electron systems where the degree of hybridization of the outermost surface layers can be reduced with respect to the bulk \cite{Laubschat1990,Suga_Ce_2000,Suga_Sm_2007}. 

The HAXPES data were measured at the beamline P22 at DESY/PETRA-III in Hamburg.\cite{Schlueter2019,Amorese2020} The experiment was performed under UHV condition of 5$\cdot$10$^{-10}$\,mbar at about $T$\,=\,60\,K on the sample with an incident photon energy of 6\,keV. The high incident energy assures that the probing depths for the Ce3$d$ core level and valence band measuremets differ only by a few percent\,\cite{TanumaIMFP_2011}. The Fermi edge of a Au foil was measured to convert from kinetic energy into binding energy. The instrumental resolution was about 200\,meV. Just before the measurements, the samples were scraped in a vacuum of 5$\cdot$10$^{-9}$\,mbar to ensure a clean surface, and the O1$s$ emission line at about 530\,eV binding energy was monitored during the experiment to verify that the sample surfaces remained clean throughout the experiment (see Fig.\,\ref{O1s} in Appendix).

The experimental data have been compared to simulations of the spectra using calculations including full multiplet (fm) and configuration interaction (CI) using the XTLS9.0 code\,\cite{Tanaka1994} for quantitative analysis of the Ce\,3$d$ emission. The fm part describes the multiplet structures of the 4$f^1$ and 4$f^2$ configurations (note the 4$f^0$ configuration consists only of one single line) and the spin orbit splitting of the 3$d_{5/2}$ and 3$d_{3/2}$ emission lines, while the four CI parameters, the 4$f$-4$f$ Coulomb repulsion ($U_{ff}$), the $4f$-$3d$ (or $3p$) core hole Coulomb attraction ($U_{4f,3d}$ or $U_{4f,3p}$), the effective hybridization $V_{eff}$, and the effective binding energy of the 4$f$ electrons $\epsilon_{4f}$, determine the energy separations and intensities of the respective 4$f$ configurations within the $3d_{3/2}$ and $3d_{5/3}$ (or $3p_{1/2}$ or $3p_{3/2}$) manifolds. Further fitting details of the fm part are provided in the Appendix. For the CI part we use the simplified model by Imer \textit{et al.} where the valence states are represented by only one ligand state\,\cite{Imer1987}, which allows the combination with the fm calculation in a way that the computational aspects are manageable. This model describes the core-level spectra very well but does not give realistic numbers for e.g.\ the Kondo temperature. The advantages and short comings of the simplified model are discussed in great detail in Ref.\,\onlinecite{Strigari2015,Sundermann2016}.

The valence band spectra were compared to band structure calculations. For this purpose density functional theory based calculations were performed using FPLO (v.18.00.52), making use of the LDA and including spin–orbit coupling in a full relativistic manner\,\cite{Koepernik1999} and weighted for the respective shell-specific photoionization cross-sections\,\cite{TRZHASKOVSKAYA2006245}. Grids of 15\,$\cdot$\,15\,$\cdot$\,15 k points and 1 energy point every 8\,meV were used for the calculation of the band structure and the density of states (DOS). The simulated valence band spectra have been obtained from the calculated cerium, transition metal and tin partial DOSs, broadened with a Gaussian lineshape with a FWHM of 0.5\,eV. Takegami \textit{et al.} have shown the validity of such analysis for hard x-ray valence band spectra\,\cite{Takegami2019}.

\section{Sample Characterization}

Figure~\ref{long}\,a shows emission scans of the CeRh$_{1-x}$Pd$_x$Sn (x\,=\,0, 0.1, 0.3, 0.5, 0.75) samples covering a large energy range that confirm the high quality of the samples. The Ce and Sn emission lines precisely fall on top of each other, only the intensities of the Rh and Pd emission lines vary due to the increasing substitution of Rh by Pd. Other, unwanted contributions are not present. The blow-up of the Pd3$s$ and Rh3$s$ emission lines between 610 and 695\,eV binding energy in Fig.\,\ref{long}\,b further confirms that the nominal composition (lines) is very much in agreement with the actual composition (dots) of the substitution series. The lines in Fig.\,\ref{long}\,b represent smoothed data of the pure Rh sample and 75\%\,Pd sample summed to the nominal amounts. %Note, that the Rh compound shows no signal in the Pd\,3$s$ region while the calculated Pd spectrum shows no signal in the Rh\,3$s$ region, confirming the 20\%\,Rh content in the 80\%\,Pd sample and the pureness of CeRhSn.

\section{Results and Analysis}

%%%%%%%%%%%%%%%%%%%%%%%%%
\begin{figure}[t]
		\includegraphics[width=0.9\columnwidth]{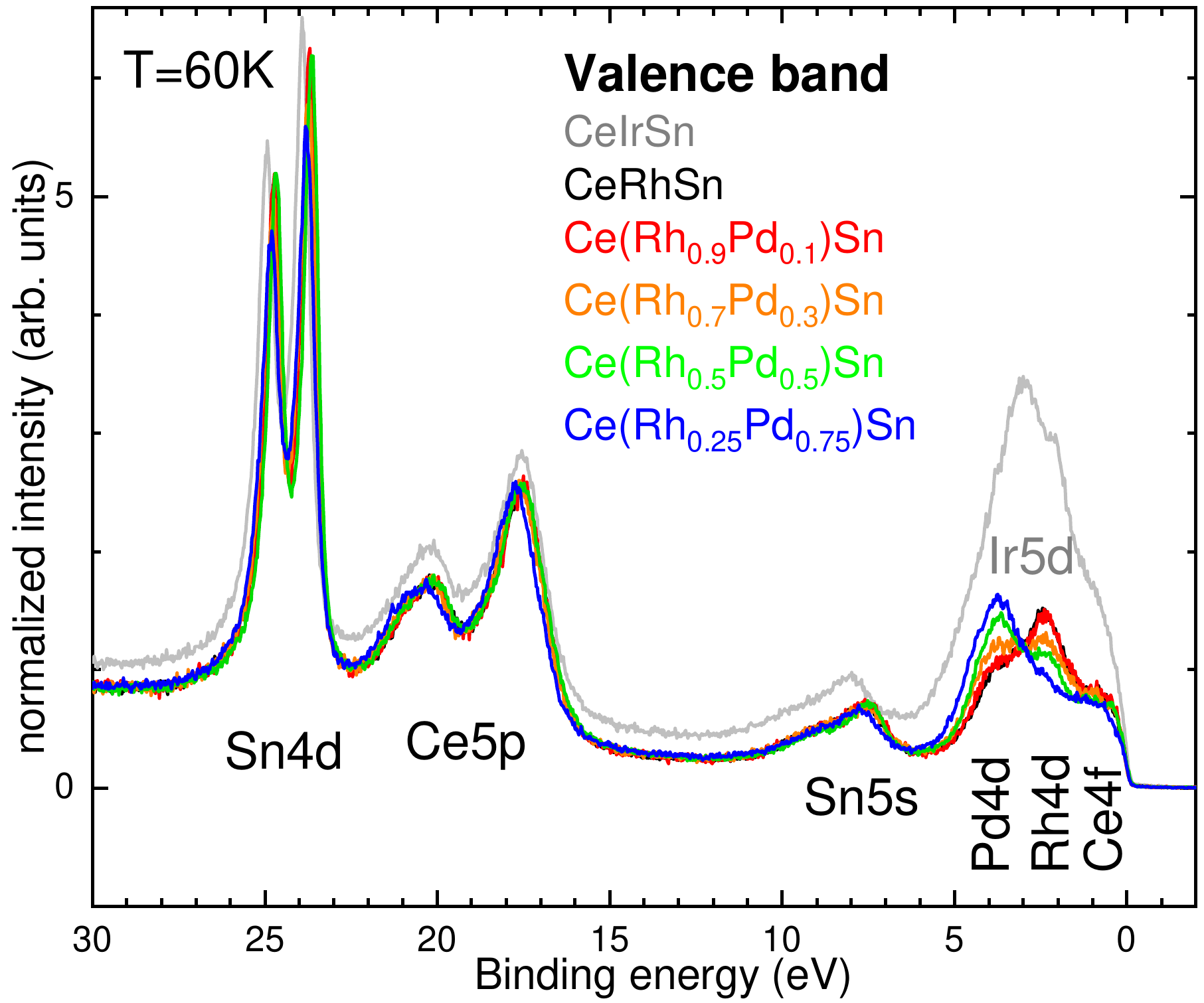} 
  \caption{(Color online) Valence band data of CeRh$_{1-x}$Pd$_x$Sn with $x$\,=\,0, 0.1, 0.3, 0.5, 0.75, and of CeIrSn at $T$\,=\,60K.} 
 \label{VB}
\end{figure}
%%%%%%%%%%%%%%%

\begin{figure*}[]
		\includegraphics[width=1.8\columnwidth]{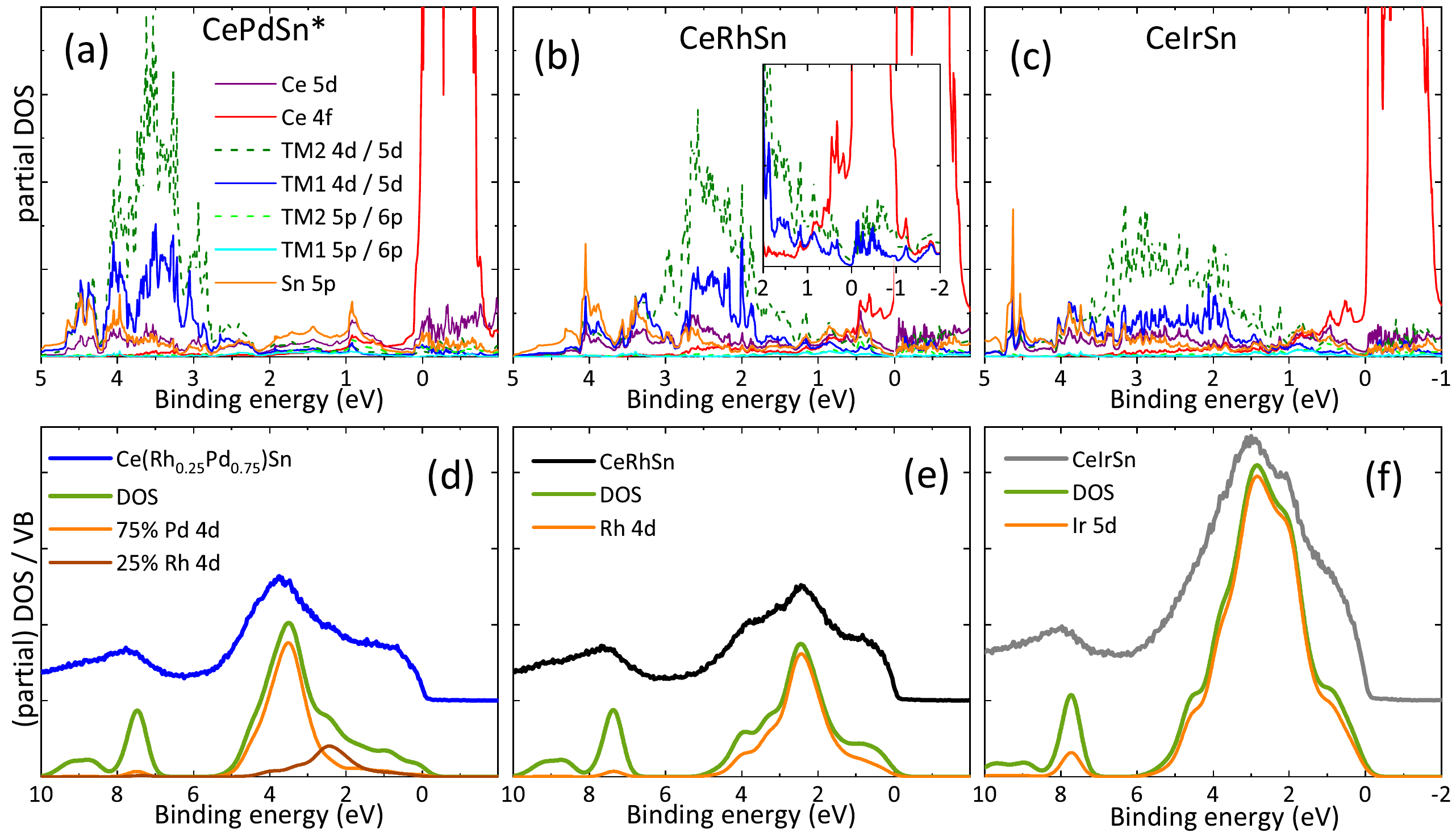} 
  \caption{(Color online) (a)-(c) Calculated partial DOS of CePdSn*, CeRhSn, and CeIrSn. CePdSn* stands for a fictitious CePdSn compound with ZrNiAl structure and lattice constants of CeRh$_{0.75}$Pd$_{0.25}$Sn. The inset in (b) provides a blow-up at the Fermi energy for Ce\,$f$ and TM\,$d$. (d)-(f) cross-section corrected and broadened calculated DOS (green lines) and partial DOS of the TM $d$ states (orange and dark orange lines) compared to valence band data of CeRh$_{0.75}$Pd$_{0.25}$Sn (blue), CeRhSn (black), and CeIrSn (grey).} 
 \label{DOS}
\end{figure*}
%%%%%%%%%%%%%%%

Figure\,\ref{Ce3d}\,a shows the Ce\,3$d$ core level of the CeRhSn and the four Pd-substituted compositions. The data are normalized to the integrated intensities across the Ce\,3$d$ core level, and a systematic trend can be seen even in the raw data; namely an increase of intensity of the main emission lines at 885 and 905\,eV and a decrease of the shoulder at 880, the bump at 895, and the satellite at 915\,eV with rising Pd concentration.

In PES, plasmon satellites may mislead the interpretation of the data. We therefore measured the Sn\,3$p$ emission lines of all CeRh$_{1-x}$Pd$_x$Sn compositions with good statistics (see Fig\,\ref{Ce3d}\,b), because the Sn spectrum is not affected by any configuration interaction effects. Hence, these spectra are good for searching for the presence of plasmons. And indeed, at 14\,eV above the main emission lines, all spectra exhibit fairly strong plasmon signals. With the Sn\,3$p$ spectra, the plasmon lineshape and energy are determined (see Appendix Table\,II). These parameters are important for the plasmon correction of the Ce emission spectra, especially when the plasmons are so close in energy to hybridization induced satellites. In addition to the plasmon correction (black lines), the Ce3$d$ spectra in Fig.\,\ref{Ce3d}\,a have to be corrected for the Shirley type background (BG) (dashed black line), and - this is special in Sn containing Ce compounds - for the intensity due to the Sn\,3$s$ emission (cyan line) at about 881\,eV. Details for the correction process are given in the Appendix.

In figure\,\ref{Ce3d}\,c, that shows the corrected spectra, the shifts of spectral weights upon substitution with Pd become even more visible. Here the respective spectral weights are indexed according to their 4$f$ occupation $f^0$, $f^1$, and $f^2$ in the initial state, and the regions of the $f^2$ spectral weight of the Ce\,3$d_{5/2}$ core level and of the 4$f^0$ of the Ce\,3$d_{3/2}$ are blown up in the insets. The strong $f^0$ intensity in CeRhSn suggests that the 4$f$ state is almost as strongly hybridized as in intermediate valent CeIrSn\,\cite{Shimura2021}. The 10\% substitution of Rh by Pd has only a minor effect on the $f$ spectral weights. Though we expect strong hybridization for 30 and 50\% substitution as well, the $f^0$ spectral weight has dropped considerably, and it has almost disappeared for 75\% Pd (see insets of Figure\,\ref{Ce3d}\,c). 

In Fig.\,\ref{Ce3d}\,c we model this trend with the fmCI calculation. We are able to capture the general trend of the spectral weight's shift very well and also the overall line shape is well reproduced which is shown separately for each Pd concentration in Fig.\,\ref{fit} in the Appendix. The resulting CI parameters are listed in Table\,I. Further details of the line-shape and fm calculation are provided in the Appendix. We have measured the Ce\,3$p$ core level (see Appendix) in addition to the Ce\,3$d$ because they are free of any other emission lines\,\cite{Sundermann2017}, and we find also that the Ce\,3$p$ spectra are well reproduced with the same CI parameters when only adjusting the Lorentzian FWHM (see Appendix Fig.\,\ref{Ce3p} and Table\,II). 

We further present valence band (VB) spectra of the CeRh$_{1-x}$Pd$_x$Sn substitution series, and also of CeIrSn for comparison, in Fig.\,\ref{VB}. The data are normalized to the Ce\,4$p$, Sn\,4$d$, and Sn\,5$s$ lines which agree very well. Only the Ir intensities are higher due to the much stronger cross-section of Ir\,5$d$ . Differences are visible at the valence band where we the expect the Pd\,4$d$ and Rh\,4$d$, and Ir\,5$d$ emission lines.  The Ce\,4$f$ (not resolved) are expected closest to the valence band.

In Fig.\,\ref{DOS}\,(a-c) we show the calculated partial density of states (DOS) at the Fermi energy of CePdSn*, CeRhSn, and also CeIrSn for comparison. The asterix in CePdSn* indicates that the DOS was calculated for a fictitious CePdSn compound with ZrNiAl structure and lattice constants of CeRh$_{0.25}$Pd$_{0.75}$Sn\,\cite{Niehaus2015}. It turns out that the transition metal (TM) $d$ states at about 3-4\,eV below the Fermi edge and the 4$f$ states at the Fermi edge are the most important contributions to the DOS in the binding energy window of 10\,eV. TM1 refers to the transition metal in the Ce plane and TM2 to the ones in the Sn plane (see Fig.\,\ref{structure}). The partial DOS of the TM1 (blue lines) and TM2 (dashed green lines)  $d$ states strongly overlap in energy reflecting that the interatomic distances d(Ce-TM1)\,=\,3.091\AA and d(Ce-TM2)\,=\,3.036\AA are not very different\cite{Niehaus2015}. The DOS of TM2, however, is about twice as strong since there are two TM2 atoms and only one TM1 in the unit cell.

Finally, we compare the VB data with the broadened and cross-section corrected DOS in Fig.\,\ref{DOS}\,(d-f) and confirm for all three compounds the validity of the DOS calculation. The calculations reproduce the same shift of spectral weight and show that it is due to the replacement of Rh\,4$d$ states by Pd\,4$d$ upon substitution, the latter being further away from the valence band.

\section{Discussion}

\begin{figure}[t]
\centering
		\includegraphics[width=0.9\columnwidth]{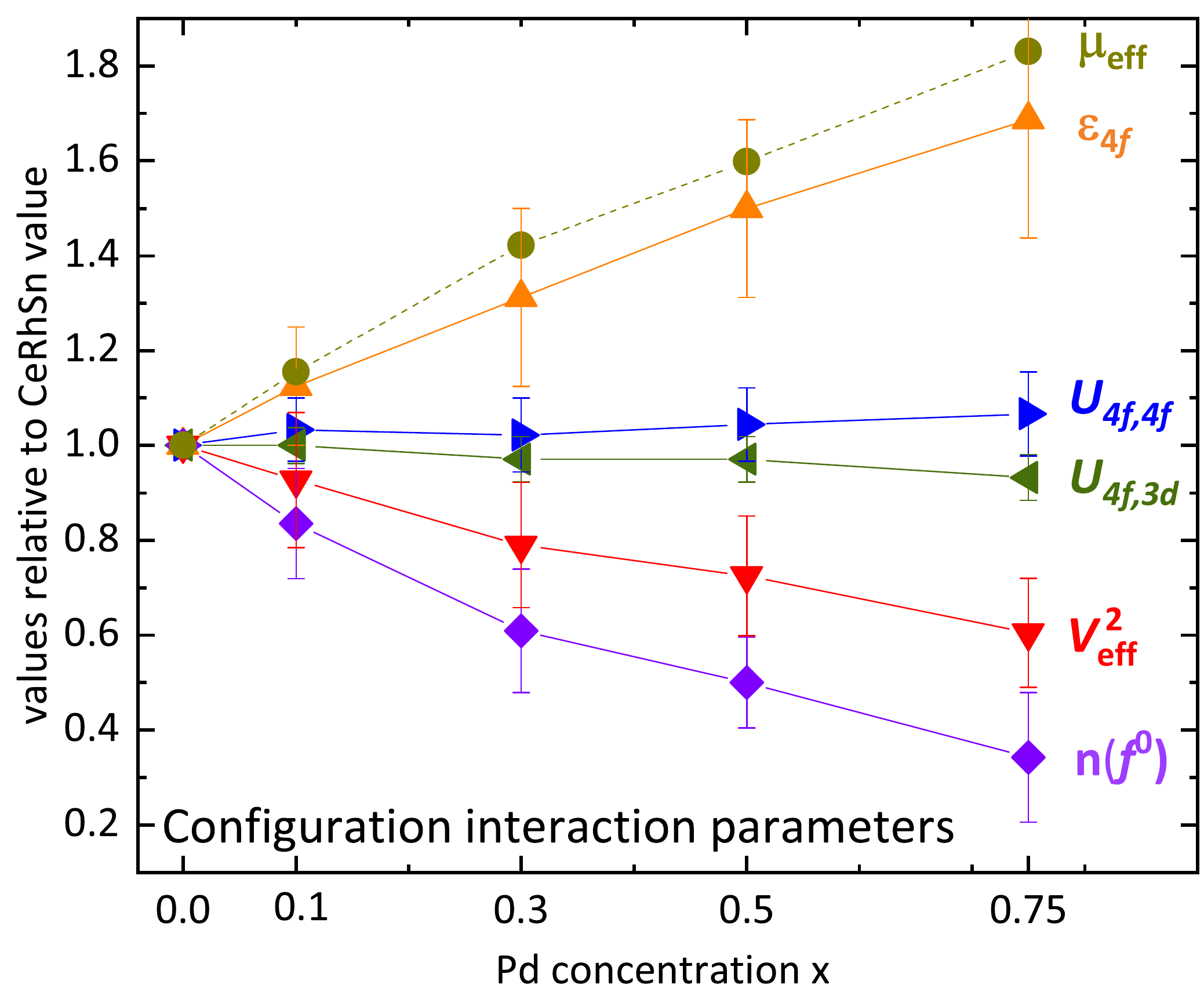} 
  \caption{(Color online) Configuration interaction parameters as function of Pd substitution $x$ normalized to the respective values of CeRhSn. The effective magnetic moments are adapted from Ref.\,\onlinecite{Yang2017}.} 
 \label{para}
\end{figure}

\begin{table*}
   \centering
    \caption{Results from fitting the corrected 3$d$ and 3$p$ HAXPES spectra of CeRh$_{1-x}$Pd$_x$Sn with the fmCI model (see Fig.\,\ref{Ce3d}\,d and Fig.\,\ref{Ce3p}). The top rows list the respective $f^i$, $i$\,=0,1,2 contributions. The corresponding CI parameters are listed in the bottom rows: the $f$-Coulomb exchange $U_{4f,4f}$, the Coulomb interaction between the $f$ electron and the 3$d$ (3$p$) core hole $U_{4f,3d}$ ($U_{4f,3p}$), the effective binding energy $\varepsilon_{4f}$, and the hybridization strength $V_\text{eff}$. The uncertainty values were estimated by varying the CI parameters independently.}
    \label{Tab_simulation}
    \begin{tabular*}{0.98\textwidth}{@{\extracolsep{\fill}}lcrrrrrr}
	\hline \hline
	                & unit  & CeRhSn    & CeRh$_{0.9}$Pd$_{0.1}$Sn & CeRh$_{0.7}$Pd$_{0.3}$Sn & CeRh$_{0.5}$Pd$_{0.5}$Sn & CeRh$_{0.25}$Pd$_{0.75}$Sn  \\
	\hline
	n($f^0$)        & \%        & 14.7\,(20) & 12.3\,(17)  &  9.0\,(19)               &  6.8\,(14)               &  4.7\,(20)                 \\
	n($f^1$)        & \%        & 82.9\,(18) & 85.4\,(16)  & 88.8\,(17)               & 91.0\,(14)               & 93.4\,(14)                  \\
	n($f^2$)        & \%        &  2.4\,(3)  &  2.3\,(3)   &  2.2\,(3)                &  2.2\,(3)                &  1.9\,(7)                  \\ 
	\\
	$U_{4f,4f}$     & eV        &  9.0\,(6)  &  9.3\,(6)   &  9.2\,(7)                &  9.4\,(7)                &  9.6\,(8)                  \\
	$U_{4f,3d}$     & eV        & 10.4\,(4)  & 10.4\,(4)   & 10.1\,(5)                & 10.1\,(5)                &  9.7\,(5)                 \\
    $U_{4f,3p}$     & eV        & 10.75\,(6)  & 10.9\,(6)   & 10.6\,(6)                & 10.5\,(9)                & 10.3\,(20)                \\
	$V_\text{eff}$  & eV        &  0.27\,(2) & 0.26\,(2)   &  0.24\,(2)               &  0.23\,(2)               &  0.21\,(2)                \\
$\varepsilon_{4f}$  & eV        &  1.6\,(2)  &  1.8\,(2)   &  2.1\,(3)                &  2.4\,(3)                & 2.7\,$^{+1.3}_{-0.4}$           \\
	\hline
    \end{tabular*}
\end{table*}

Figure\,\ref{para} summarizes the findings of the the fmCI analysis of the core-level data for CeRh$_{1-x}$Pd$_x$Sn. The CI parameters, the amount of n($f^0$), the effective magnetic moments (taken from Ref.\,\,\onlinecite{Yang2017}) and unit cell volume\,\cite{Niehaus2015} (not shown) all change monotonically with increasing Pd concentration $x$. Note, all values in Fig.\,\ref{para} are normalized to the respective values of CeRhSn. The parameters $U_{4f,4f}$ and $U_{4f,3d}$ do not vary much across the CeRh$_{1-x}$Pd$_x$Sn series, but the hybridization $V_\text{eff}$ (red) decreases and the effective 4$f$ binding energy $\varepsilon_{4f}$ (orange) increases with increasing Pd concentration. Both the smaller $V_\text{eff}$ and the larger $\varepsilon_{4f}$ have the effect of stabilizing the $f^1$ configuration because it becomes energetically more expensive for the $f$ electron to escape into the conduction band. Hence n($f^0$) (purple) decreases.  This is very much in agreement with the recovery of the effective magnetic moments (dark yellow) observed in magnetic measurements\,\cite{Yang2017}.  The valences obtained from the present quantitative analysis differ from the numbers given in Ref.\,\onlinecite{Niehaus2015} because Niehaus \textit{et al.}\,did not apply a configuration interaction calculation, hence ignoring final state effects. The PES data of CeRhSn in Ref.\,\onlinecite{Gamza2009} find less 4$f^0$ in CeRhSn which could be due to the stronger surface sensitivity of PES with respect to HAXPES. In general, however, it is found that the Rh rich side of the phase diagram of CeRh$_{1-x}$Sn$_x$ is strongly intermediate valent ($\alpha$-Ce like). In fact, the 4$f$ electrons in CeRhSn are almost as strongly hybridized as in CeIrSn\,\cite{Shimura2021}. The Pd rich side, on the other hand, is more like a $\gamma$-type Ce compound\cite{Murani1993}. 

The present valence band data as well as the calculation of the partial  DOS of CeRhSn agree very well with the data and calculations by Gamza \textit{et al.}\cite{Gamza2009}. In addition, we find that the Pd4$d$ states in CePdSn* are about 1\,eV further away from the Fermi edge than the Rh4$d$ and Ir5$d$ states in CeRhSn and CeIrSn, respectively. Hence, the energy difference to the Ce4$f$ is larger by this amount. This is very much in agreement with the CI results that find an increase of the effective binding energy of the Ce\,4$f$ electrons, $\varepsilon_{4f}$, by about the same energy when going from CeRhSn to CeRh$_{0.25}$Pd$_{0.75}$Sn (see Table\,I) or with respect to CeIrSn\,\cite{Shimura2021}. Accordingly, the hybridization in the Pd rich samples is weaker than in CeRhSn and also weaker than in CeIrSn. It is amazing that these two very different approaches for data analysis, i.e. the Anderson impurity calculation for the simulation of the core level data and the band structure calculation for reproducing the valence band data yield the same shift of energy scale as function of the Pd substitution $x$. Both, however, the larger $\epsilon_{4f}$ and the larger energy separation of TM4$d$ and Ce4$f$ states go along with reduced hybridization. From this we draw the conclusion that the TM\,$d$ states are driving the hybridization with the Ce\,4$f$ states. %The DFT calculations provide 4$f$ shell fillings 1.124, 1.0923, and 1.092 for CePdSn*, CeRhSn, and CeIrSn, respectively, i.e.\ the DOS calculations provide the correct trend (see Table\,I) despite some overestimation.

We can only speculate which one of the two Rh atoms in CeRhSn is most important for hybridization effects. According to the partial DOS both Rh1 and Rh2 overlap with the Ce\,4$f$ states in the vicinity of the Fermi energy (see inset in Fig.\,\ref{DOS}\,b). Kittaka \textit{et al.} suggest that a $J_z$\,=\,$\pm$3/2 ground state is compatible with the strongly anisotropic non-Fermi liquid behavior that they have observed\,\cite{Kittaka2021}. Here the $J_z$\,=\,$\pm$3/2 state refers to the yo-yo shaped 4$f$ charge density, which should favor hybridization with the Rh2 atoms that are in the Sn and not in the Ce plane (see Fig.\,\ref{structure}).
We should not forget, however, that CeRhSn is strongly intermediate valent and that Kondo temperature and crystal-field splittings, based on neutron scattering data of CePtSn and CePdSn\,\cite{Kohgi1993}, are comparable. Hence, it is questionable whether these symmetry aspects can be applied to CeRhSn.

K{\"u}chler \textit{et al.} and Yang \textit{et al.}\,\cite{Kuchler2017,Yang2017} both argued that in CeRhSn signatures of local moments exist and that CeRhSn under uniaxial pressure is close to a quantum critical point, very much like CeIrSn at ambient pressure\,\cite{Tsuda2018}. Furthermore, in both compounds metamagnetic cross-overs occur under the field along the $a$ axis, and in CeIrSn antiferromagentic correlations have been observed at very low temperature\,\cite{Shimura2021} thus feeding the speculation that magnetic order may be hindered by frustration despite the strong hybridization of 4$f$ and conduction electrons. The present quantitative analysis of the 4$f$ electronic configurations in CeRhSn shows that also CeRhSn belongs to the strongly intermediate valence regime, where, usually, magnetic order does not occur, thus, making the proximity to a quantum critical point very interesting. We believe that it is a very complex task to disentangle the impact of frustration and hybridization in such strongly hybridized compounds as CeRhSn and CeIrSn. This is very much in contrast to the low $T_K$ system CePdAl where frustration was unambiguously proven\,\cite{Donni1996,Oyamada2008,Fritsch2014,Fritsch2017,Lucas2017} and seems to persist on substituting Pd with Ni up to the quantum critical point in CePd$_{1-x}$Ni$_x$Al at $x$\,$\approx$\,0.135\,\cite{Huesges2017}. In the CeRh$_{1-x}$Pd$_x$Sn series, however, we found that 4$f$-conduction electron hybridization is still quite pronounced for $x$\,=\,0.1 and also for $x$\,=\,0.3 so that also here magnetic order should be hindered by the Kondo interaction. The origin of the maxima in both $C/T$ and $\chi_{ac}(T)$ remains therefore an open puzzle. 

Below 100\,K domains of intermediate valent Ce (80\%) and Ce$^{3+}$ are reported from $\mu$SR measurements\,\cite{Schenck2004}. We consider this to be unlikely in view of the smooth change of the CI parameters with the Pd concentration $x$ in CeRh$_{1-x}$Pd$_x$Sn at 60\,K and the good agreement of the energy scales obtained from the Anderson impurity analysis of the core level spectra and the \textit{ab initio} calculation of the partial DOS that describe the valence band data, unless the same fraction of domains persists throughout the substitution series. 

\section{Summary}
The substitution series of the quasikagome Kondo compounds CeRh$_{1-x}$Pd$_x$Sn with $x$\,=\,0, 0.1, 0.3, 0.5, 0.75 has been investigated with core-level and valence band HAXPES. Both the pure Rh compound and the sample substituted with 10\,\% Pd belong to the intermediate valent regime. Effects of $cf$-hybridization are strongly suppressed for the higher Pd concentrations. The quantitative configuration interaction and full multiplet analysis show that the substitution of Pd leads to a smooth decrease of the hybridization $V_\text{eff}$ and an increase of the effective 4$f$ binding energy $\varepsilon_{4f}$. We also find that the binding energy increases for the transition metal $d$ states in the valence band data upon substitution of Pd and this increase scales with the effective binding energy of 4$f$ electrons $\varepsilon_{4f}$. This suggests that the transition metal 4$d$ states drive the hybridization.

\section{Appendix}

%%%%%%%%%%%%%%%%%%%%%%%%%
\begin{figure}[]
 \centering
		\includegraphics[width=0.98\columnwidth]{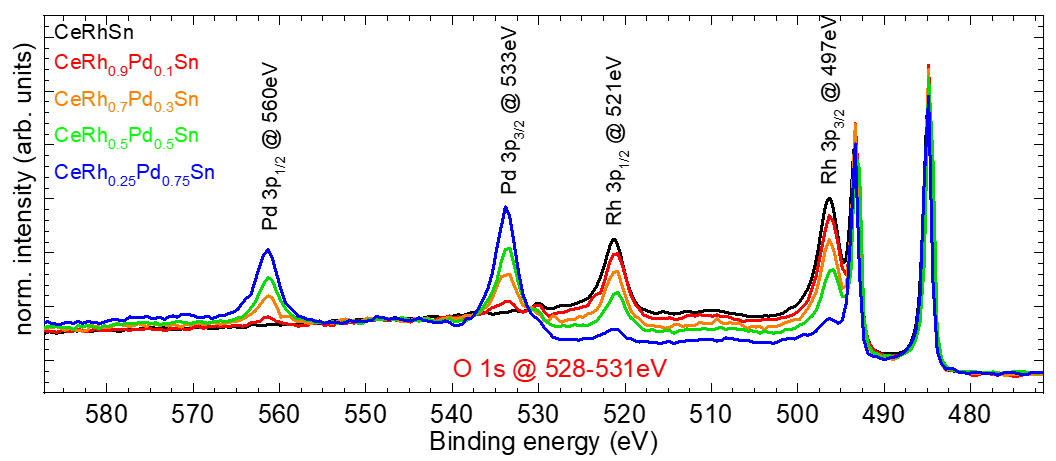} 
  \caption{(Color online) Photo-electron emission scans of CeRh$_{1-x}$Pd$_x$Sn with $x$\,=\,0, 0.1, 0.3, 0.5, 0.75 over the energy range of O1s. The emission energies for the Pd3p and Rh3p lines were taken from Ref.\,\cite{Handbook} using the energies 
for Pd and Rh metals, and for the O1s using the listed energies for metal oxides.} 
 \label{O1s}
\end{figure}
%%%%%%%%%%%%%%%

%%%%%%%%%%%%%%%%%%%%%%%%%
\begin{figure*}[]
 \centering
		\includegraphics[width=2.1\columnwidth]{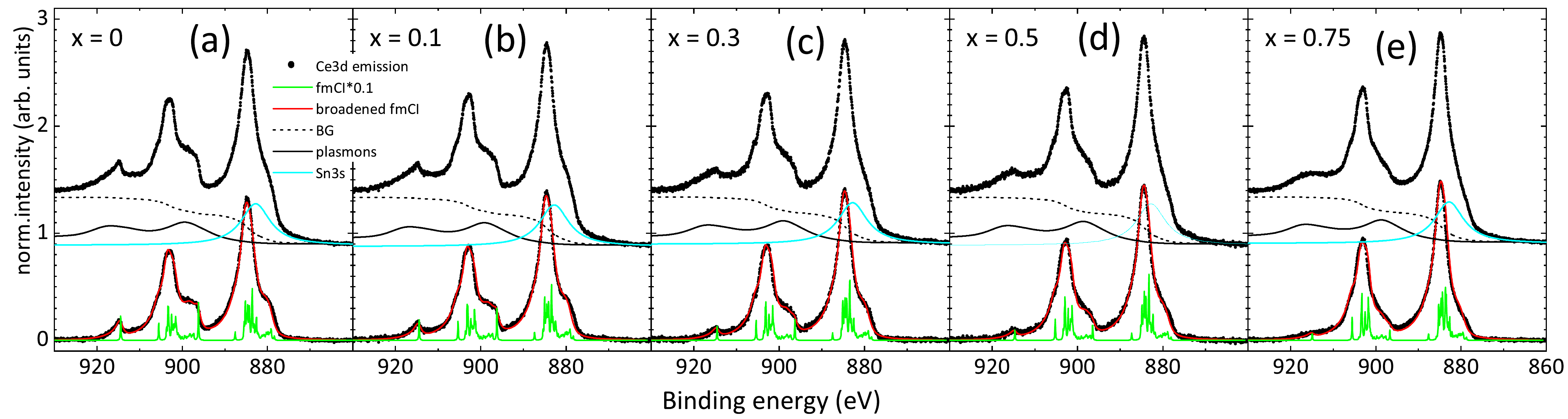} 
  \caption{(Color online) Ce\,3$d$ core-level emission lines of CeRh$_{1-x}$Pd$_x$Sn with x\,=\,0, 0.1, 0.3, 0.5, 0.75 (a-e). At the top the experimental data are shown along with the corresponding integrated background (BG) and the plasmon and Sn\,3$s$ contributions. Below, the Ce\,3$d$ emission spectra obtained from subtracting BG, plasmon, and Sn\,3$s$ from the data are shown along with the broadened full multiple (fm) configuration interaction (CI) calculation (red line). A less broadened fmCI calculation times 0.1 (green line) illustrates the underlaying multiplet lines.} 
 \label{fit}
\end{figure*}
%%%%%%%%%%%%%%%

\begin{table}[]
    \caption{Lineshape and plasmon parameters: Lorentzian FWHM$_L$ and Gaussian full widths at half maximum FWHM$_G$, the Mahan assymetry factor $\alpha_M$ and cutoff energy $\gamma_M$, and the intensity relative to the emission line $A^{Pl}$ and widths of the plamsons FWHM$_L^{Pl}$. Different plamsons energies $\Delta\text{E}^{Pl}$ of 14.8, 14.7, 14.5, 14.3, 14.0\,eV have been obtained form the Sn\,3$p$ spectra for $x$\,=\,0, 0.1, 0.3, 0.5, 0.75, respectively.}
    \label{Tab_simulation}
  \begin{tabular*}{0.48\textwidth}{@{\extracolsep{\fill}}lcrrr}
	\hline
	                & unit   & 3$d$       & 3$p$      \\
	\hline
	FWHM$_L$        & eV     & 1.3        & 5.0         \\
	FWHM$_G$        & eV     & \multicolumn{2}{c}{0.3} \\
	$\alpha_M$      & 1      & \multicolumn{2}{c}{0.4} \\ 
	$\gamma_M$      & eV     & \multicolumn{2}{c}{4}   \\
	$A^{Pl}$        & 1      & \multicolumn{2}{c}{0.24}\\
    FWHM$_L^{Pl}$   & eV     & \multicolumn{2}{c}{6}  \\
	\hline
    \end{tabular*}
\end{table}

In the fm part of the calculation the atomic values of the intra-atomic $4f$-$4f$ and $3d$-$4f$ ($3p$-$4f$) Coulomb and exchange interactions and the 3$d$ (3$p$) and 4$f$ spin-orbit coupling are taken from the Cowan code. Best agreement with the experimental core-level emission structure is obtained when the $4f$-$4f$ ($3d$-$4f$ and $3p$-$4f$) Coulomb interactions are reduced to 60\% (80\%) and the spin-orbit coupling of the 3$d$ (3$p$) is reduced to 98\% (97.5\%). The line shape of the main emission lines is described in terms of a Voigt profile (Gaussian FWHM$_G$ and Lorentzian width FWHM$_L$) convoluted with Mahan asymmetry $\alpha_M$ and cutoff energy $\gamma_M$.\,\cite{Mahan1975} These lineshape parameters are listed in Table\,II. They have been optimized once and were then kept constant for all Pd concentration.

For a better comparison with the fmCI calculations, an integrated background and the plasmon contribution, as obtained from the Sn\,3$p$ emission has been removed from the Ce core-level emission lines. Figure\,\ref{fit}(a-e) show the deconvolution of the CeRh$_{1-x}$Pd$_x$Sn with x\,=\,0, 0.1, 0.3, 0.5, 0.75 core-level spectra. The black dashed line is the integrated Shirley-type background (BG), the back solid line describes the plasmons with the parameters given in Table\,II as obtained from the Sn\,3$p$ spectra in Fig.\,\ref{long}\,(c). The light blue line represents the Sn\,3$s$ spectral weight which has been determined iteratively. The spectrum resulting from the core-level spectra after subtracting the three contributions (BG, plasmon, Sn\,3$s$) reproduce the 6/8 intensity ratio of the Ce\,3$d_{3/2}$ and Ce\,3$d_{5/2}$ spectral weights.
% and thus give us confidence that the correction including the Sn\,3$s$ spectral weight has been performed properly.
The corrected Ce\,3$d$ emission lines are in good agreement with the Ce\,3$d$ simulations.

%%%%%%%%%%%%%%%%%%%%%%%%%
\begin{figure}[]
 \centering
		\includegraphics[width=0.9\columnwidth]{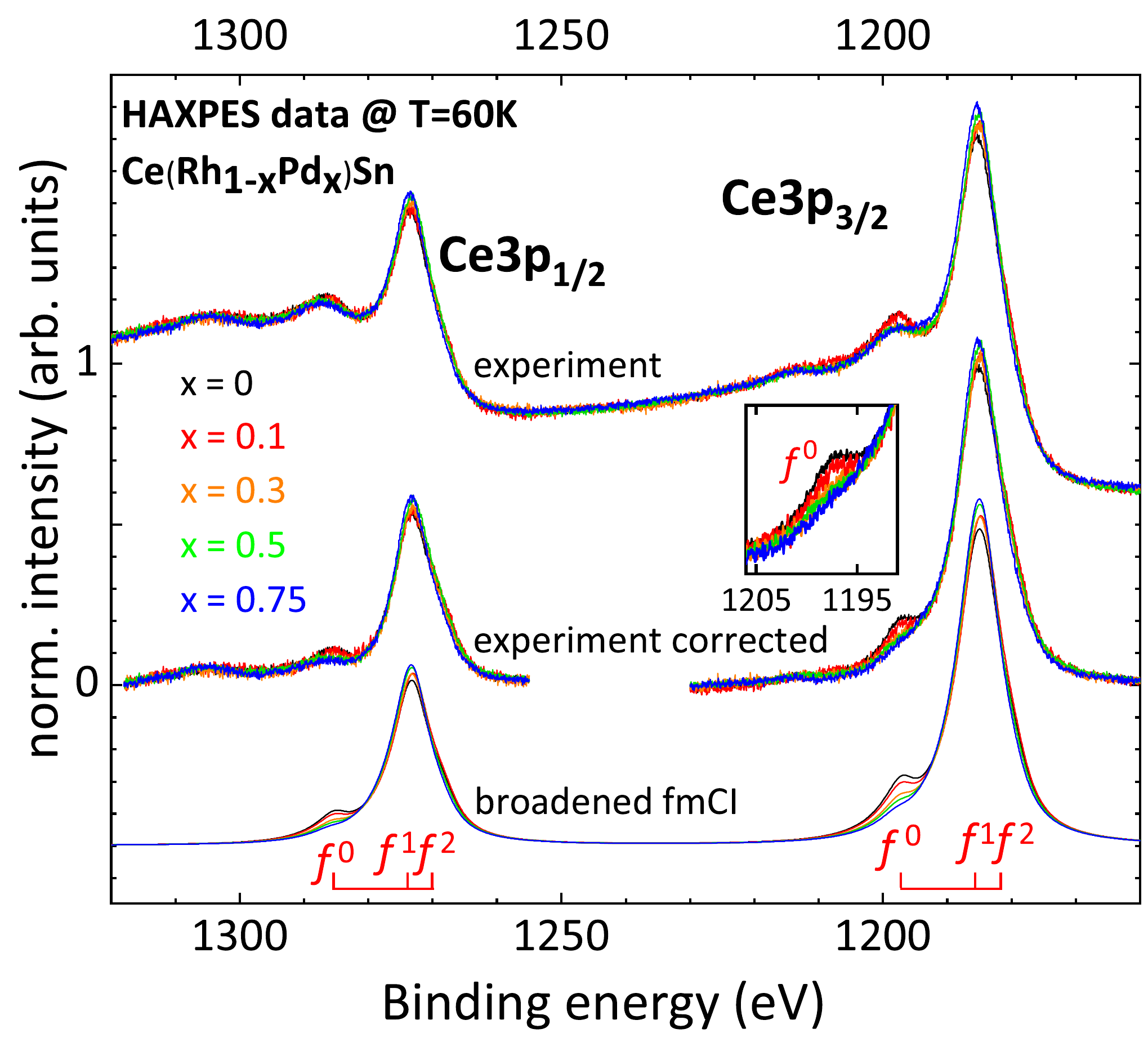} 
  \caption{(Color online)  Ce\,3$p$ core-level emission spectra and analysis of CeRh$_{1-x}$Pd$_x$Sn with $x$\,=\,0, 0.1, 0.3, 0.5, 0.75. From top to bottom, the experimental data as measured, the corrected data after BG and plasmon subtraction, and the fmCI simulations with same CI parameters as for Ce\,3$d$ are shown. The inset is a blow up of the $f^0$ region of the Ce\,3$p_{3/2}$. } 
 \label{Ce3p}
\end{figure}
%%%%%%%%%%%%%%%

In Sn containing compounds, the photoelectron spectroscopy spectra of the Ce\,3$d$ emission lines have the ambiguity of the correction for the Sn\,3$s$ intensity beneath the Ce\,3$d_{5/2}$. We have therefore measured as well the Ce\,3$p$ emission lines (see Fig.\,\ref{Ce3p}) and analyzed them in the same manner as the Ce\,3$d$ emission lines. The Ce\,3$p_{1/2}$ and Ce\,3$p_{3/2}$ do not overlap with any other emission line. They are further apart in energy and broader than the Ce\,3$d$. The $f^0$ intensities are still well separated from the $f^1$ spectral weights but the $f^1$ and $f^2$ spectral weights are not resolved. Nevertheless, also the Ce\,3$p$ spectra show the same shifts of spectral weights as function of Pd substitution, and most importantly, this trend can be captured with same parameters as for the Ce\,3$d$. Only the line broadening had to be adjusted accounting for a different core-hole life time (see Table II). 

\section{Acknowledgement}
We would like to thank D.\,T Adroja for many fruitful discussions. This work was supported by projects JPSJ KAKENHI No. 21K03473 and No. 17K05545 and by the German Research Foundation (DFG) under project SE 1441-5-2. We further acknowledge DESY (Hamburg, Germany), a member of the Helmholtz Association HGF, for the provision of experimental facilities.


\begin{thebibliography}{59}%
\makeatletter
\providecommand \@ifxundefined [1]{%
 \@ifx{#1\undefined}
}%
\providecommand \@ifnum [1]{%
 \ifnum #1\expandafter \@firstoftwo
 \else \expandafter \@secondoftwo
 \fi
}%
\providecommand \@ifx [1]{%
 \ifx #1\expandafter \@firstoftwo
 \else \expandafter \@secondoftwo
 \fi
}%
\providecommand \natexlab [1]{#1}%
\providecommand \enquote  [1]{``#1''}%
\providecommand \bibnamefont  [1]{#1}%
\providecommand \bibfnamefont [1]{#1}%
\providecommand \citenamefont [1]{#1}%
\providecommand \href@noop [0]{\@secondoftwo}%
\providecommand \href [0]{\begingroup \@sanitize@url \@href}%
\providecommand \@href[1]{\@@startlink{#1}\@@href}%
\providecommand \@@href[1]{\endgroup#1\@@endlink}%
\providecommand \@sanitize@url [0]{\catcode `\\12\catcode `\$12\catcode
  `\&12\catcode `\#12\catcode `\^12\catcode `\_12\catcode `\%12\relax}%
\providecommand \@@startlink[1]{}%
\providecommand \@@endlink[0]{}%
\providecommand \url  [0]{\begingroup\@sanitize@url \@url }%
\providecommand \@url [1]{\endgroup\@href {#1}{\urlprefix }}%
\providecommand \urlprefix  [0]{URL }%
\providecommand \Eprint [0]{\href }%
\providecommand \doibase [0]{https://doi.org/}%
\providecommand \selectlanguage [0]{\@gobble}%
\providecommand \bibinfo  [0]{\@secondoftwo}%
\providecommand \bibfield  [0]{\@secondoftwo}%
\providecommand \translation [1]{[#1]}%
\providecommand \BibitemOpen [0]{}%
\providecommand \bibitemStop [0]{}%
\providecommand \bibitemNoStop [0]{.\EOS\space}%
\providecommand \EOS [0]{\spacefactor3000\relax}%
\providecommand \BibitemShut  [1]{\csname bibitem#1\endcsname}%
\let\auto@bib@innerbib\@empty
%</preamble>
\bibitem [{\citenamefont {Coleman}(2007)}]{Coleman2007}%
  \BibitemOpen
  \bibfield  {author} {\bibinfo {author} {\bibfnamefont {P.}~\bibnamefont
  {Coleman}},\ }\href@noop {} {\emph {\bibinfo {title} {{H}andbook of {M}agn.
  and {A}dv. {M}agn. {M}ater.}}},\ edited by\ \bibinfo {editor} {\bibfnamefont
  {M.~F. S.~M.}\ \bibnamefont {H.~Kronm\"uller}, \bibfnamefont {S.~Parkin}}\
  and\ \bibinfo {editor} {\bibfnamefont {I.}~\bibnamefont {Zutic}},\
  Vol.~\bibinfo {volume} {1}\ (\bibinfo  {publisher} {John Wiley and Sons},\
  \bibinfo {year} {2007})\ pp.\ \bibinfo {pages} {95--148},\ \bibinfo {note}
  {"Heavy fermions: electrons at the edge of magnetism"}\BibitemShut {NoStop}%
\bibitem [{\citenamefont {Coleman}(2015)}]{Coleman2015}%
  \BibitemOpen
  \bibfield  {author} {\bibinfo {author} {\bibfnamefont {P.}~\bibnamefont
  {Coleman}},\ }\bibfield  {title} {\bibinfo {title} {{Heavy fermions and the
  Kondo lattice: A 21st century perspective}},\ }\href@noop {} {\bibfield
  {journal} {\bibinfo  {journal} {edt. E. Pavarini, E. Koch, and P. Coleman}\
  }\textbf {\bibinfo {volume} {(Forschungszentrum J\"ulich)}},\ \bibinfo
  {pages} {95} (\bibinfo {year} {2015})},\ \bibinfo {note} {{in Many-Body
  Physics: From Kondo to Hubbard}}\BibitemShut {NoStop}%
\bibitem [{\citenamefont {Doniach}(1977)}]{Doniach1977}%
  \BibitemOpen
  \bibfield  {author} {\bibinfo {author} {\bibfnamefont {S.}~\bibnamefont
  {Doniach}},\ }\bibfield  {title} {\bibinfo {title} {The {K}ondo lattice and
  weak antiferromagnetism},\ }\href@noop {} {\bibfield  {journal} {\bibinfo
  {journal} {Physica B}\ }\textbf {\bibinfo {volume} {91}},\ \bibinfo {pages}
  {231} (\bibinfo {year} {1977})}\BibitemShut {NoStop}%
\bibitem [{\citenamefont {L\"ohneysen}\ \emph {et~al.}(2007)\citenamefont
  {L\"ohneysen}, \citenamefont {Rosch}, \citenamefont {Vojta},\ and\
  \citenamefont {W\"olfle}}]{Hilbert2007}%
  \BibitemOpen
  \bibfield  {author} {\bibinfo {author} {\bibfnamefont {H.~v.}\ \bibnamefont
  {L\"ohneysen}}, \bibinfo {author} {\bibfnamefont {A.}~\bibnamefont {Rosch}},
  \bibinfo {author} {\bibfnamefont {M.}~\bibnamefont {Vojta}},\ and\ \bibinfo
  {author} {\bibfnamefont {P.}~\bibnamefont {W\"olfle}},\ }\bibfield  {title}
  {\bibinfo {title} {Fermi-liquid instabilities at magnetic quantum phase
  transitions},\ }\href {https://doi.org/10.1103/RevModPhys.79.1015} {\bibfield
   {journal} {\bibinfo  {journal} {Rev. Mod. Phys.}\ }\textbf {\bibinfo
  {volume} {79}},\ \bibinfo {pages} {1015} (\bibinfo {year}
  {2007})}\BibitemShut {NoStop}%
\bibitem [{\citenamefont {Wirth}\ and\ \citenamefont
  {Steglich}(2016)}]{Wirth2016}%
  \BibitemOpen
  \bibfield  {author} {\bibinfo {author} {\bibfnamefont {S.}~\bibnamefont
  {Wirth}}\ and\ \bibinfo {author} {\bibfnamefont {F.}~\bibnamefont
  {Steglich}},\ }\bibfield  {title} {\bibinfo {title} {Exploring heavy fermions
  from macroscopic to microscopic length scales},\ }\href
  {https://doi.org/10.1038/natrevmats.2016.51} {\bibfield  {journal} {\bibinfo
  {journal} {Nat. Rev. Mater.}\ }\textbf {\bibinfo {volume} {1}},\ \bibinfo
  {pages} {16066} (\bibinfo {year} {2016})}\BibitemShut {NoStop}%
\bibitem [{\citenamefont {Grigera}\ \emph {et~al.}(2004)\citenamefont
  {Grigera}, \citenamefont {Gegenwart}, \citenamefont {Borzi}, \citenamefont
  {Weickert}, \citenamefont {Schofield}, \citenamefont {Perry}, \citenamefont
  {Tayama}, \citenamefont {Sakakibara}, \citenamefont {Maeno}, \citenamefont
  {Green},\ and\ \citenamefont {Mackenzie}}]{Grigera2004}%
  \BibitemOpen
  \bibfield  {author} {\bibinfo {author} {\bibfnamefont {S.~A.}\ \bibnamefont
  {Grigera}}, \bibinfo {author} {\bibfnamefont {P.}~\bibnamefont {Gegenwart}},
  \bibinfo {author} {\bibfnamefont {R.~A.}\ \bibnamefont {Borzi}}, \bibinfo
  {author} {\bibfnamefont {F.}~\bibnamefont {Weickert}}, \bibinfo {author}
  {\bibfnamefont {A.~J.}\ \bibnamefont {Schofield}}, \bibinfo {author}
  {\bibfnamefont {R.~S.}\ \bibnamefont {Perry}}, \bibinfo {author}
  {\bibfnamefont {T.}~\bibnamefont {Tayama}}, \bibinfo {author} {\bibfnamefont
  {T.}~\bibnamefont {Sakakibara}}, \bibinfo {author} {\bibfnamefont
  {Y.}~\bibnamefont {Maeno}}, \bibinfo {author} {\bibfnamefont {A.~G.}\
  \bibnamefont {Green}},\ and\ \bibinfo {author} {\bibfnamefont {A.~P.}\
  \bibnamefont {Mackenzie}},\ }\bibfield  {title} {\bibinfo {title}
  {Disorder-sensitive phase formation linked to metamagnetic quantum
  criticality},\ }\href {https://doi.org/10.1126/science.1104306} {\bibfield
  {journal} {\bibinfo  {journal} {Science}\ }\textbf {\bibinfo {volume}
  {306}},\ \bibinfo {pages} {1154} (\bibinfo {year} {2004})}\BibitemShut
  {NoStop}%
\bibitem [{\citenamefont {Senthil}\ \emph {et~al.}(2004)\citenamefont
  {Senthil}, \citenamefont {Vojta},\ and\ \citenamefont
  {Sachdev}}]{Senthil2004}%
  \BibitemOpen
  \bibfield  {author} {\bibinfo {author} {\bibfnamefont {T.}~\bibnamefont
  {Senthil}}, \bibinfo {author} {\bibfnamefont {M.}~\bibnamefont {Vojta}},\
  and\ \bibinfo {author} {\bibfnamefont {S.}~\bibnamefont {Sachdev}},\
  }\bibfield  {title} {\bibinfo {title} {Weak magnetism and non-{F}ermi liquids
  near heavy-fermion critical points},\ }\href
  {https://doi.org/10.1103/PhysRevB.69.035111} {\bibfield  {journal} {\bibinfo
  {journal} {Phys. Rev. B}\ }\textbf {\bibinfo {volume} {69}},\ \bibinfo
  {pages} {035111} (\bibinfo {year} {2004})}\BibitemShut {NoStop}%
\bibitem [{\citenamefont {Si}(2006)}]{Si2006}%
  \BibitemOpen
  \bibfield  {author} {\bibinfo {author} {\bibfnamefont {Q.}~\bibnamefont
  {Si}},\ }\bibfield  {title} {\bibinfo {title} {Global magnetic phase diagram
  and local quantum criticality in heavy fermion metals},\ }\href
  {https://doi.org/https://doi.org/10.1016/j.physb.2006.01.156} {\bibfield
  {journal} {\bibinfo  {journal} {Physica B: Cond. Matter}\ }\textbf {\bibinfo
  {volume} {378-380}},\ \bibinfo {pages} {23} (\bibinfo {year} {2006})},\
  \bibinfo {note} {proceedings of the International Conference on Strongly
  Correlated Electron Systems}\BibitemShut {NoStop}%
\bibitem [{\citenamefont {Vojta}(2008)}]{Vojta2010}%
  \BibitemOpen
  \bibfield  {author} {\bibinfo {author} {\bibfnamefont {M.}~\bibnamefont
  {Vojta}},\ }\bibfield  {title} {\bibinfo {title} {From itinerant to
  local-moment antiferromagnetism in kondo lattices: Adiabatic continuity
  versus quantum phase transitions},\ }\href
  {https://doi.org/10.1103/PhysRevB.78.125109} {\bibfield  {journal} {\bibinfo
  {journal} {Phys. Rev. B}\ }\textbf {\bibinfo {volume} {78}},\ \bibinfo
  {pages} {125109} (\bibinfo {year} {2008})}\BibitemShut {NoStop}%
\bibitem [{\citenamefont {Coleman}\ and\ \citenamefont
  {Nevidomskyy}(2010)}]{Coleman2010}%
  \BibitemOpen
  \bibfield  {author} {\bibinfo {author} {\bibfnamefont {P.}~\bibnamefont
  {Coleman}}\ and\ \bibinfo {author} {\bibfnamefont {A.~H.}\ \bibnamefont
  {Nevidomskyy}},\ }\bibfield  {title} {\bibinfo {title} {Frustration and the
  {K}ondo effect in heavy fermion materials},\ }\href
  {https://doi.org/10.1007/s10909-010-0213-4} {\bibfield  {journal} {\bibinfo
  {journal} {J. Low Temp. Phys.}\ }\textbf {\bibinfo {volume} {161}},\ \bibinfo
  {pages} {182} (\bibinfo {year} {2010})}\BibitemShut {NoStop}%
\bibitem [{\citenamefont {Si}\ \emph {et~al.}(2014)\citenamefont {Si},
  \citenamefont {Pixley}, \citenamefont {Nica}, \citenamefont {Yamamoto},
  \citenamefont {Goswami}, \citenamefont {Yu},\ and\ \citenamefont
  {Kirchner}}]{Si2014}%
  \BibitemOpen
  \bibfield  {author} {\bibinfo {author} {\bibfnamefont {Q.}~\bibnamefont
  {Si}}, \bibinfo {author} {\bibfnamefont {J.~H.}\ \bibnamefont {Pixley}},
  \bibinfo {author} {\bibfnamefont {E.}~\bibnamefont {Nica}}, \bibinfo {author}
  {\bibfnamefont {S.~J.}\ \bibnamefont {Yamamoto}}, \bibinfo {author}
  {\bibfnamefont {P.}~\bibnamefont {Goswami}}, \bibinfo {author} {\bibfnamefont
  {R.}~\bibnamefont {Yu}},\ and\ \bibinfo {author} {\bibfnamefont
  {S.}~\bibnamefont {Kirchner}},\ }\bibfield  {title} {\bibinfo {title}
  {{K}ondo destruction and quantum criticality in {K}ondo lattice systems},\
  }\href {https://doi.org/10.7566/JPSJ.83.061005} {\bibfield  {journal}
  {\bibinfo  {journal} {J. Phys. Soc. Jpn.}\ }\textbf {\bibinfo {volume}
  {83}},\ \bibinfo {pages} {061005} (\bibinfo {year} {2014})}\BibitemShut
  {NoStop}%
\bibitem [{\citenamefont {Lacroix}(2010)}]{Lacroix2010}%
  \BibitemOpen
  \bibfield  {author} {\bibinfo {author} {\bibfnamefont {C.}~\bibnamefont
  {Lacroix}},\ }\bibfield  {title} {\bibinfo {title} {Frustrated metallic
  systems: A review of some peculiar behavior},\ }\href
  {https://doi.org/10.1143/JPSJ.79.011008} {\bibfield  {journal} {\bibinfo
  {journal} {J. Phys. Soc. Jpn.}\ }\textbf {\bibinfo {volume} {79}},\ \bibinfo
  {pages} {011008} (\bibinfo {year} {2010})}\BibitemShut {NoStop}%
\bibitem [{\citenamefont {D\"onni}\ \emph {et~al.}(1996)\citenamefont
  {D\"onni}, \citenamefont {Ehlers}, \citenamefont {Maletta}, \citenamefont
  {Fischer}, \citenamefont {Kitazawa},\ and\ \citenamefont
  {Zolliker}}]{Donni1996}%
  \BibitemOpen
  \bibfield  {author} {\bibinfo {author} {\bibfnamefont {A.}~\bibnamefont
  {D\"onni}}, \bibinfo {author} {\bibfnamefont {G.}~\bibnamefont {Ehlers}},
  \bibinfo {author} {\bibfnamefont {H.}~\bibnamefont {Maletta}}, \bibinfo
  {author} {\bibfnamefont {P.}~\bibnamefont {Fischer}}, \bibinfo {author}
  {\bibfnamefont {H.}~\bibnamefont {Kitazawa}},\ and\ \bibinfo {author}
  {\bibfnamefont {M.}~\bibnamefont {Zolliker}},\ }\bibfield  {title} {\bibinfo
  {title} {Geometrically frustrated magnetic structures of the heavy-fermion
  compound {CePdAl} studied by powder neutron diffraction},\ }\href
  {https://doi.org/10.1088/0953-8984/8/50/043} {\bibfield  {journal} {\bibinfo
  {journal} {J. Phys.: Cond. Matt.}\ }\textbf {\bibinfo {volume} {8}},\
  \bibinfo {pages} {11213} (\bibinfo {year} {1996})}\BibitemShut {NoStop}%
\bibitem [{\citenamefont {Oyamada}\ \emph {et~al.}(2008)\citenamefont
  {Oyamada}, \citenamefont {Maegawa}, \citenamefont {Nishiyama}, \citenamefont
  {Kitazawa},\ and\ \citenamefont {Isikawa}}]{Oyamada2008}%
  \BibitemOpen
  \bibfield  {author} {\bibinfo {author} {\bibfnamefont {A.}~\bibnamefont
  {Oyamada}}, \bibinfo {author} {\bibfnamefont {S.}~\bibnamefont {Maegawa}},
  \bibinfo {author} {\bibfnamefont {M.}~\bibnamefont {Nishiyama}}, \bibinfo
  {author} {\bibfnamefont {H.}~\bibnamefont {Kitazawa}},\ and\ \bibinfo
  {author} {\bibfnamefont {Y.}~\bibnamefont {Isikawa}},\ }\bibfield  {title}
  {\bibinfo {title} {Ordering mechanism and spin fluctuations in a
  geometrically frustrated heavy-fermion antiferromagnet on the {K}agome-like
  lattice {CePdAl}: A $^{27}${Al} {NMR} study},\ }\href
  {https://doi.org/10.1103/PhysRevB.77.064432} {\bibfield  {journal} {\bibinfo
  {journal} {Phys. Rev. B}\ }\textbf {\bibinfo {volume} {77}},\ \bibinfo
  {pages} {064432} (\bibinfo {year} {2008})}\BibitemShut {NoStop}%
\bibitem [{\citenamefont {Fritsch}\ \emph {et~al.}(2014)\citenamefont
  {Fritsch}, \citenamefont {Bagrets}, \citenamefont {Goll}, \citenamefont
  {Kittler}, \citenamefont {Wolf}, \citenamefont {Grube}, \citenamefont
  {Huang},\ and\ \citenamefont {L\"ohneysen}}]{Fritsch2014}%
  \BibitemOpen
  \bibfield  {author} {\bibinfo {author} {\bibfnamefont {V.}~\bibnamefont
  {Fritsch}}, \bibinfo {author} {\bibfnamefont {N.}~\bibnamefont {Bagrets}},
  \bibinfo {author} {\bibfnamefont {G.}~\bibnamefont {Goll}}, \bibinfo {author}
  {\bibfnamefont {W.}~\bibnamefont {Kittler}}, \bibinfo {author} {\bibfnamefont
  {M.~J.}\ \bibnamefont {Wolf}}, \bibinfo {author} {\bibfnamefont
  {K.}~\bibnamefont {Grube}}, \bibinfo {author} {\bibfnamefont {C.-L.}\
  \bibnamefont {Huang}},\ and\ \bibinfo {author} {\bibfnamefont {H.~v.}\
  \bibnamefont {L\"ohneysen}},\ }\bibfield  {title} {\bibinfo {title}
  {Approaching quantum criticality in a partially geometrically frustrated
  heavy-fermion metal},\ }\href {https://doi.org/10.1103/PhysRevB.89.054416}
  {\bibfield  {journal} {\bibinfo  {journal} {Phys. Rev. B}\ }\textbf {\bibinfo
  {volume} {89}},\ \bibinfo {pages} {054416} (\bibinfo {year}
  {2014})}\BibitemShut {NoStop}%
\bibitem [{\citenamefont {Fritsch}\ \emph {et~al.}(2017)\citenamefont
  {Fritsch}, \citenamefont {Lucas}, \citenamefont {Huesges}, \citenamefont
  {Sakai}, \citenamefont {Kittler}, \citenamefont {Taubenheim}, \citenamefont
  {Woitschach}, \citenamefont {Pedersen}, \citenamefont {Grube}, \citenamefont
  {Schmidt}, \citenamefont {Gegenwart}, \citenamefont {Stockert},\ and\
  \citenamefont {v.~L\"ohneysen}}]{Fritsch2017}%
  \BibitemOpen
  \bibfield  {author} {\bibinfo {author} {\bibfnamefont {V.}~\bibnamefont
  {Fritsch}}, \bibinfo {author} {\bibfnamefont {S.}~\bibnamefont {Lucas}},
  \bibinfo {author} {\bibfnamefont {Z.}~\bibnamefont {Huesges}}, \bibinfo
  {author} {\bibfnamefont {A.}~\bibnamefont {Sakai}}, \bibinfo {author}
  {\bibfnamefont {W.}~\bibnamefont {Kittler}}, \bibinfo {author} {\bibfnamefont
  {C.}~\bibnamefont {Taubenheim}}, \bibinfo {author} {\bibfnamefont
  {S.}~\bibnamefont {Woitschach}}, \bibinfo {author} {\bibfnamefont
  {B.}~\bibnamefont {Pedersen}}, \bibinfo {author} {\bibfnamefont
  {K.}~\bibnamefont {Grube}}, \bibinfo {author} {\bibfnamefont
  {B.}~\bibnamefont {Schmidt}}, \bibinfo {author} {\bibfnamefont
  {P.}~\bibnamefont {Gegenwart}}, \bibinfo {author} {\bibfnamefont
  {O.}~\bibnamefont {Stockert}},\ and\ \bibinfo {author} {\bibfnamefont
  {H.}~\bibnamefont {v.~L\"ohneysen}},\ }\bibfield  {title} {\bibinfo {title}
  {{CePdAl} - a {K}ondo lattice with partial frustration},\ }\href
  {https://doi.org/10.1088/1742-6596/807/3/032003} {\bibfield  {journal}
  {\bibinfo  {journal} {J. Phys.: Conf. Series}\ }\textbf {\bibinfo {volume}
  {807}},\ \bibinfo {pages} {032003} (\bibinfo {year} {2017})}\BibitemShut
  {NoStop}%
\bibitem [{\citenamefont {Lucas}\ \emph {et~al.}(2017)\citenamefont {Lucas},
  \citenamefont {Grube}, \citenamefont {Huang}, \citenamefont {Sakai},
  \citenamefont {Wunderlich}, \citenamefont {Green}, \citenamefont {Wosnitza},
  \citenamefont {Fritsch}, \citenamefont {Gegenwart}, \citenamefont
  {Stockert},\ and\ \citenamefont {v.~L\"ohneysen}}]{Lucas2017}%
  \BibitemOpen
  \bibfield  {author} {\bibinfo {author} {\bibfnamefont {S.}~\bibnamefont
  {Lucas}}, \bibinfo {author} {\bibfnamefont {K.}~\bibnamefont {Grube}},
  \bibinfo {author} {\bibfnamefont {C.-L.}\ \bibnamefont {Huang}}, \bibinfo
  {author} {\bibfnamefont {A.}~\bibnamefont {Sakai}}, \bibinfo {author}
  {\bibfnamefont {S.}~\bibnamefont {Wunderlich}}, \bibinfo {author}
  {\bibfnamefont {E.~L.}\ \bibnamefont {Green}}, \bibinfo {author}
  {\bibfnamefont {J.}~\bibnamefont {Wosnitza}}, \bibinfo {author}
  {\bibfnamefont {V.}~\bibnamefont {Fritsch}}, \bibinfo {author} {\bibfnamefont
  {P.}~\bibnamefont {Gegenwart}}, \bibinfo {author} {\bibfnamefont
  {O.}~\bibnamefont {Stockert}},\ and\ \bibinfo {author} {\bibfnamefont
  {H.}~\bibnamefont {v.~L\"ohneysen}},\ }\bibfield  {title} {\bibinfo {title}
  {Entropy evolution in the magnetic phases of partially frustrated {CePdAl}},\
  }\href {https://doi.org/10.1103/PhysRevLett.118.107204} {\bibfield  {journal}
  {\bibinfo  {journal} {Phys. Rev. Lett.}\ }\textbf {\bibinfo {volume} {118}},\
  \bibinfo {pages} {107204} (\bibinfo {year} {2017})}\BibitemShut {NoStop}%
\bibitem [{\citenamefont {Huesges}\ \emph {et~al.}(2017)\citenamefont
  {Huesges}, \citenamefont {Lucas}, \citenamefont {Wunderlich}, \citenamefont
  {Yokaichiya}, \citenamefont {Proke\ifmmode~\check{s}\else \v{s}\fi{}},
  \citenamefont {Schmalzl}, \citenamefont {Lem\'ee-Cailleau}, \citenamefont
  {Pedersen}, \citenamefont {Fritsch}, \citenamefont {v.~L\"ohneysen},\ and\
  \citenamefont {Stockert}}]{Huesges2017}%
  \BibitemOpen
  \bibfield  {author} {\bibinfo {author} {\bibfnamefont {Z.}~\bibnamefont
  {Huesges}}, \bibinfo {author} {\bibfnamefont {S.}~\bibnamefont {Lucas}},
  \bibinfo {author} {\bibfnamefont {S.}~\bibnamefont {Wunderlich}}, \bibinfo
  {author} {\bibfnamefont {F.}~\bibnamefont {Yokaichiya}}, \bibinfo {author}
  {\bibfnamefont {K.}~\bibnamefont {Proke\ifmmode~\check{s}\else \v{s}\fi{}}},
  \bibinfo {author} {\bibfnamefont {K.}~\bibnamefont {Schmalzl}}, \bibinfo
  {author} {\bibfnamefont {M.-H.}\ \bibnamefont {Lem\'ee-Cailleau}}, \bibinfo
  {author} {\bibfnamefont {B.}~\bibnamefont {Pedersen}}, \bibinfo {author}
  {\bibfnamefont {V.}~\bibnamefont {Fritsch}}, \bibinfo {author} {\bibfnamefont
  {H.}~\bibnamefont {v.~L\"ohneysen}},\ and\ \bibinfo {author} {\bibfnamefont
  {O.}~\bibnamefont {Stockert}},\ }\bibfield  {title} {\bibinfo {title}
  {Evolution of the partially frustrated magnetic order in
  {CePd}$_{1-x}${Ni}$_x${Al}},\ }\href
  {https://doi.org/10.1103/PhysRevB.96.144405} {\bibfield  {journal} {\bibinfo
  {journal} {Phys. Rev. B}\ }\textbf {\bibinfo {volume} {96}},\ \bibinfo
  {pages} {144405} (\bibinfo {year} {2017})}\BibitemShut {NoStop}%
\bibitem [{\citenamefont {Zhao}\ \emph {et~al.}(2019)\citenamefont {Zhao},
  \citenamefont {Zhang}, \citenamefont {Lyu}, \citenamefont {Bachus},
  \citenamefont {Tokiwa}, \citenamefont {Gegenwart}, \citenamefont {Zhang},
  \citenamefont {Cheng}, \citenamefont {Yang}, \citenamefont {Chen},
  \citenamefont {Isikawa}, \citenamefont {Si}, \citenamefont {Steglich},\ and\
  \citenamefont {Sun}}]{Zhao2019}%
  \BibitemOpen
  \bibfield  {author} {\bibinfo {author} {\bibfnamefont {H.}~\bibnamefont
  {Zhao}}, \bibinfo {author} {\bibfnamefont {J.}~\bibnamefont {Zhang}},
  \bibinfo {author} {\bibfnamefont {M.}~\bibnamefont {Lyu}}, \bibinfo {author}
  {\bibfnamefont {S.}~\bibnamefont {Bachus}}, \bibinfo {author} {\bibfnamefont
  {Y.}~\bibnamefont {Tokiwa}}, \bibinfo {author} {\bibfnamefont
  {P.}~\bibnamefont {Gegenwart}}, \bibinfo {author} {\bibfnamefont
  {S.}~\bibnamefont {Zhang}}, \bibinfo {author} {\bibfnamefont
  {J.}~\bibnamefont {Cheng}}, \bibinfo {author} {\bibfnamefont {Y.-f.}\
  \bibnamefont {Yang}}, \bibinfo {author} {\bibfnamefont {G.}~\bibnamefont
  {Chen}}, \bibinfo {author} {\bibfnamefont {Y.}~\bibnamefont {Isikawa}},
  \bibinfo {author} {\bibfnamefont {Q.}~\bibnamefont {Si}}, \bibinfo {author}
  {\bibfnamefont {F.}~\bibnamefont {Steglich}},\ and\ \bibinfo {author}
  {\bibfnamefont {P.}~\bibnamefont {Sun}},\ }\bibfield  {title} {\bibinfo
  {title} {Quantum-critical phase from frustrated magnetism in a strongly
  correlated metal},\ }\href {https://doi.org/10.1038/s41567-019-0666-6}
  {\bibfield  {journal} {\bibinfo  {journal} {Nat. Phys.}\ }\textbf {\bibinfo
  {volume} {15}},\ \bibinfo {pages} {1261} (\bibinfo {year}
  {2019})}\BibitemShut {NoStop}%
\bibitem [{\citenamefont {P\"ottgen}\ and\ \citenamefont
  {Chevalier}(2015)}]{Pottgen2015}%
  \BibitemOpen
  \bibfield  {author} {\bibinfo {author} {\bibfnamefont {R.}~\bibnamefont
  {P\"ottgen}}\ and\ \bibinfo {author} {\bibfnamefont {B.}~\bibnamefont
  {Chevalier}},\ }\bibfield  {title} {\bibinfo {title} {Cerium intermetallics
  with {ZrNiAl}-type structure - a review},\ }\href
  {https://doi.org/doi:10.1515/znb-2015-0018} {\bibfield  {journal} {\bibinfo
  {journal} {Z. Naturforschung B}\ }\textbf {\bibinfo {volume} {70}},\ \bibinfo
  {pages} {289} (\bibinfo {year} {2015})}\BibitemShut {NoStop}%
\bibitem [{\citenamefont {Nohara}\ \emph {et~al.}(1993)\citenamefont {Nohara},
  \citenamefont {Namatame}, \citenamefont {Fujimori},\ and\ \citenamefont
  {Takabatake}}]{Nohara1993}%
  \BibitemOpen
  \bibfield  {author} {\bibinfo {author} {\bibfnamefont {S.}~\bibnamefont
  {Nohara}}, \bibinfo {author} {\bibfnamefont {H.}~\bibnamefont {Namatame}},
  \bibinfo {author} {\bibfnamefont {A.}~\bibnamefont {Fujimori}},\ and\
  \bibinfo {author} {\bibfnamefont {T.}~\bibnamefont {Takabatake}},\ }\bibfield
   {title} {\bibinfo {title} {Photoemission study of {CeNiSn} and related
  compounds},\ }\href {https://doi.org/10.1103/PhysRevB.47.1754} {\bibfield
  {journal} {\bibinfo  {journal} {Phys. Rev. B}\ }\textbf {\bibinfo {volume}
  {47}},\ \bibinfo {pages} {1754} (\bibinfo {year} {1993})}\BibitemShut
  {NoStop}%
\bibitem [{\citenamefont {Kim}\ \emph {et~al.}(2003)\citenamefont {Kim},
  \citenamefont {Echizen}, \citenamefont {Umeo}, \citenamefont {Kobayashi},
  \citenamefont {Sera}, \citenamefont {Salamkha}, \citenamefont {Sologub},
  \citenamefont {Takabatake}, \citenamefont {Chen}, \citenamefont {Tayama},
  \citenamefont {Sakakibara}, \citenamefont {Jung},\ and\ \citenamefont
  {Maple}}]{Kim2003}%
  \BibitemOpen
  \bibfield  {author} {\bibinfo {author} {\bibfnamefont {M.~S.}\ \bibnamefont
  {Kim}}, \bibinfo {author} {\bibfnamefont {Y.}~\bibnamefont {Echizen}},
  \bibinfo {author} {\bibfnamefont {K.}~\bibnamefont {Umeo}}, \bibinfo {author}
  {\bibfnamefont {S.}~\bibnamefont {Kobayashi}}, \bibinfo {author}
  {\bibfnamefont {M.}~\bibnamefont {Sera}}, \bibinfo {author} {\bibfnamefont
  {P.~S.}\ \bibnamefont {Salamkha}}, \bibinfo {author} {\bibfnamefont {O.~L.}\
  \bibnamefont {Sologub}}, \bibinfo {author} {\bibfnamefont {T.}~\bibnamefont
  {Takabatake}}, \bibinfo {author} {\bibfnamefont {X.}~\bibnamefont {Chen}},
  \bibinfo {author} {\bibfnamefont {T.}~\bibnamefont {Tayama}}, \bibinfo
  {author} {\bibfnamefont {T.}~\bibnamefont {Sakakibara}}, \bibinfo {author}
  {\bibfnamefont {M.~H.}\ \bibnamefont {Jung}},\ and\ \bibinfo {author}
  {\bibfnamefont {M.~B.}\ \bibnamefont {Maple}},\ }\bibfield  {title} {\bibinfo
  {title} {Low-temperature anomalies in magnetic, transport, and thermal
  properties of single-crystal {CeRhSn} with valence fluctuations},\ }\href
  {https://doi.org/10.1103/PhysRevB.68.054416} {\bibfield  {journal} {\bibinfo
  {journal} {Phys. Rev. B}\ }\textbf {\bibinfo {volume} {68}},\ \bibinfo
  {pages} {054416} (\bibinfo {year} {2003})}\BibitemShut {NoStop}%
\bibitem [{\citenamefont {Schenck}\ \emph {et~al.}(2004)\citenamefont
  {Schenck}, \citenamefont {Gygax}, \citenamefont {Kim},\ and\ \citenamefont
  {Takabatake}}]{Schenck2004}%
  \BibitemOpen
  \bibfield  {author} {\bibinfo {author} {\bibfnamefont {A.}~\bibnamefont
  {Schenck}}, \bibinfo {author} {\bibfnamefont {F.~N.}\ \bibnamefont {Gygax}},
  \bibinfo {author} {\bibfnamefont {M.~S.}\ \bibnamefont {Kim}},\ and\ \bibinfo
  {author} {\bibfnamefont {T.}~\bibnamefont {Takabatake}},\ }\bibfield  {title}
  {\bibinfo {title} {Study of the electronic properties of {CeRhSn} by $\mu ^+$
  knight shift and relaxation measurements in single crystals},\ }\href
  {https://doi.org/10.1143/JPSJ.73.3099} {\bibfield  {journal} {\bibinfo
  {journal} {J. Phys. Soc. Jpn.}\ }\textbf {\bibinfo {volume} {73}},\ \bibinfo
  {pages} {3099} (\bibinfo {year} {2004})}\BibitemShut {NoStop}%
\bibitem [{\citenamefont {Tokiwa}\ \emph {et~al.}(2015)\citenamefont {Tokiwa},
  \citenamefont {Stingl}, \citenamefont {Kim}, \citenamefont {Takabatake},\
  and\ \citenamefont {Gegenwart}}]{Tokiwae2015}%
  \BibitemOpen
  \bibfield  {author} {\bibinfo {author} {\bibfnamefont {Y.}~\bibnamefont
  {Tokiwa}}, \bibinfo {author} {\bibfnamefont {C.}~\bibnamefont {Stingl}},
  \bibinfo {author} {\bibfnamefont {M.-S.}\ \bibnamefont {Kim}}, \bibinfo
  {author} {\bibfnamefont {T.}~\bibnamefont {Takabatake}},\ and\ \bibinfo
  {author} {\bibfnamefont {P.}~\bibnamefont {Gegenwart}},\ }\bibfield  {title}
  {\bibinfo {title} {Characteristic signatures of quantum criticality driven by
  geometrical frustration},\ }\bibfield  {journal} {\bibinfo  {journal} {Sci.
  Adv.}\ }\textbf {\bibinfo {volume} {1}},\ \href
  {https://doi.org/10.1126/sciadv.1500001} {10.1126/sciadv.1500001} (\bibinfo
  {year} {2015})\BibitemShut {NoStop}%
\bibitem [{\citenamefont {Yang}\ \emph {et~al.}(2017)\citenamefont {Yang},
  \citenamefont {Tsuda}, \citenamefont {Umeo}, \citenamefont {Yamane},
  \citenamefont {Onimaru}, \citenamefont {Takabatake}, \citenamefont
  {Kikugawa}, \citenamefont {Terashima},\ and\ \citenamefont {Uji}}]{Yang2017}%
  \BibitemOpen
  \bibfield  {author} {\bibinfo {author} {\bibfnamefont {C.~L.}\ \bibnamefont
  {Yang}}, \bibinfo {author} {\bibfnamefont {S.}~\bibnamefont {Tsuda}},
  \bibinfo {author} {\bibfnamefont {K.}~\bibnamefont {Umeo}}, \bibinfo {author}
  {\bibfnamefont {Y.}~\bibnamefont {Yamane}}, \bibinfo {author} {\bibfnamefont
  {T.}~\bibnamefont {Onimaru}}, \bibinfo {author} {\bibfnamefont
  {T.}~\bibnamefont {Takabatake}}, \bibinfo {author} {\bibfnamefont
  {N.}~\bibnamefont {Kikugawa}}, \bibinfo {author} {\bibfnamefont
  {T.}~\bibnamefont {Terashima}},\ and\ \bibinfo {author} {\bibfnamefont
  {S.}~\bibnamefont {Uji}},\ }\bibfield  {title} {\bibinfo {title} {Quantum
  criticality and development of antiferromagnetic order in the quasikagome
  {K}ondo lattice {CeRh}$_{1-x}${P}d$_x${Sn}},\ }\href
  {https://doi.org/10.1103/PhysRevB.96.045139} {\bibfield  {journal} {\bibinfo
  {journal} {Phys. Rev. B}\ }\textbf {\bibinfo {volume} {96}},\ \bibinfo
  {pages} {045139} (\bibinfo {year} {2017})}\BibitemShut {NoStop}%
\bibitem [{\citenamefont {K\"uchler}\ \emph {et~al.}(2017)\citenamefont
  {K\"uchler}, \citenamefont {Stingl}, \citenamefont {Tokiwa}, \citenamefont
  {Kim}, \citenamefont {Takabatake},\ and\ \citenamefont
  {Gegenwart}}]{Kuchler2017}%
  \BibitemOpen
  \bibfield  {author} {\bibinfo {author} {\bibfnamefont {R.}~\bibnamefont
  {K\"uchler}}, \bibinfo {author} {\bibfnamefont {C.}~\bibnamefont {Stingl}},
  \bibinfo {author} {\bibfnamefont {Y.}~\bibnamefont {Tokiwa}}, \bibinfo
  {author} {\bibfnamefont {M.~S.}\ \bibnamefont {Kim}}, \bibinfo {author}
  {\bibfnamefont {T.}~\bibnamefont {Takabatake}},\ and\ \bibinfo {author}
  {\bibfnamefont {P.}~\bibnamefont {Gegenwart}},\ }\bibfield  {title} {\bibinfo
  {title} {Uniaxial stress tuning of geometrical frustration in a {K}ondo
  lattice},\ }\href {https://doi.org/10.1103/PhysRevB.96.241110} {\bibfield
  {journal} {\bibinfo  {journal} {Phys. Rev. B}\ }\textbf {\bibinfo {volume}
  {96}},\ \bibinfo {pages} {241110(R)} (\bibinfo {year} {2017})}\BibitemShut
  {NoStop}%
\bibitem [{\citenamefont {Grosche}\ \emph {et~al.}(2001)\citenamefont
  {Grosche}, \citenamefont {Walker}, \citenamefont {Julian}, \citenamefont
  {Mathur}, \citenamefont {Freye}, \citenamefont {Steiner},\ and\ \citenamefont
  {Lonzarich}}]{Grosche2001}%
  \BibitemOpen
  \bibfield  {author} {\bibinfo {author} {\bibfnamefont {F.~M.}\ \bibnamefont
  {Grosche}}, \bibinfo {author} {\bibfnamefont {I.~R.}\ \bibnamefont {Walker}},
  \bibinfo {author} {\bibfnamefont {S.~R.}\ \bibnamefont {Julian}}, \bibinfo
  {author} {\bibfnamefont {N.~D.}\ \bibnamefont {Mathur}}, \bibinfo {author}
  {\bibfnamefont {D.~M.}\ \bibnamefont {Freye}}, \bibinfo {author}
  {\bibfnamefont {M.~J.}\ \bibnamefont {Steiner}},\ and\ \bibinfo {author}
  {\bibfnamefont {G.~G.}\ \bibnamefont {Lonzarich}},\ }\bibfield  {title}
  {\bibinfo {title} {Superconductivity on the threshold of magnetism in
  {CePd}$_2${S}i$_2$ and {CeIn}$_3$},\ }\href
  {https://doi.org/10.1088/0953-8984/13/12/309} {\bibfield  {journal} {\bibinfo
   {journal} {J. Phys.: Cond. Matter}\ }\textbf {\bibinfo {volume} {13}},\
  \bibinfo {pages} {2845} (\bibinfo {year} {2001})}\BibitemShut {NoStop}%
\bibitem [{\citenamefont {Grube}\ \emph {et~al.}(2018)\citenamefont {Grube},
  \citenamefont {Pintschovius}, \citenamefont {Weber}, \citenamefont
  {Castellan}, \citenamefont {Zaum}, \citenamefont {Kuntz}, \citenamefont
  {Schweiss}, \citenamefont {Stockert}, \citenamefont {Bachus}, \citenamefont
  {Shimura}, \citenamefont {Fritsch},\ and\ \citenamefont
  {L\"ohneysen}}]{Grube2018}%
  \BibitemOpen
  \bibfield  {author} {\bibinfo {author} {\bibfnamefont {K.}~\bibnamefont
  {Grube}}, \bibinfo {author} {\bibfnamefont {L.}~\bibnamefont {Pintschovius}},
  \bibinfo {author} {\bibfnamefont {F.}~\bibnamefont {Weber}}, \bibinfo
  {author} {\bibfnamefont {J.-P.}\ \bibnamefont {Castellan}}, \bibinfo {author}
  {\bibfnamefont {S.}~\bibnamefont {Zaum}}, \bibinfo {author} {\bibfnamefont
  {S.}~\bibnamefont {Kuntz}}, \bibinfo {author} {\bibfnamefont
  {P.}~\bibnamefont {Schweiss}}, \bibinfo {author} {\bibfnamefont
  {O.}~\bibnamefont {Stockert}}, \bibinfo {author} {\bibfnamefont
  {S.}~\bibnamefont {Bachus}}, \bibinfo {author} {\bibfnamefont
  {Y.}~\bibnamefont {Shimura}}, \bibinfo {author} {\bibfnamefont
  {V.}~\bibnamefont {Fritsch}},\ and\ \bibinfo {author} {\bibfnamefont {H.~v.}\
  \bibnamefont {L\"ohneysen}},\ }\bibfield  {title} {\bibinfo {title} {Magnetic
  and structural quantum phase transitions in {CeCu}$_{6-x}${Au}$_x$ are
  independent},\ }\href {https://doi.org/10.1103/PhysRevLett.121.087203}
  {\bibfield  {journal} {\bibinfo  {journal} {Phys. Rev. Lett.}\ }\textbf
  {\bibinfo {volume} {121}},\ \bibinfo {pages} {087203} (\bibinfo {year}
  {2018})}\BibitemShut {NoStop}%
\bibitem [{\citenamefont {Niehaus}\ \emph {et~al.}(2015)\citenamefont
  {Niehaus}, \citenamefont {Abdala},\ and\ \citenamefont
  {P\"ottgen}}]{Niehaus2015}%
  \BibitemOpen
  \bibfield  {author} {\bibinfo {author} {\bibfnamefont {O.}~\bibnamefont
  {Niehaus}}, \bibinfo {author} {\bibfnamefont {P.~M.}\ \bibnamefont
  {Abdala}},\ and\ \bibinfo {author} {\bibfnamefont {R.}~\bibnamefont
  {P\"ottgen}},\ }\bibfield  {title} {\bibinfo {title} {The solid solutions
  {CeR}u$_{1-x}${P}d$_x${S}n and {CeR}h$_{1-x}${P}d$_x${S}n - {A}pplicability
  of the {ICF} model to determine intermediate cerium valencies by comparison
  with {XANES} data},\ }\href {https://doi.org/doi:10.1515/znb-2015-0003}
  {\bibfield  {journal} {\bibinfo  {journal} {Z. Naturforschung B}\ }\textbf
  {\bibinfo {volume} {70}},\ \bibinfo {pages} {253} (\bibinfo {year}
  {2015})}\BibitemShut {NoStop}%
\bibitem [{\citenamefont {Adroja~$et$ $al$.}()}]{Adroja}%
  \BibitemOpen
  \bibfield  {author} {\bibinfo {author} {\bibfnamefont {D.}~\bibnamefont
  {Adroja~$et$ $al$.}},\ }\bibfield  {title} {\bibinfo {title} {work in
  progress}\ }\href@noop {} {}\BibitemShut {NoStop}%
\bibitem [{\citenamefont {Gunnarsson}\ \emph {et~al.}(2001)\citenamefont
  {Gunnarsson}, \citenamefont {Sch\"onhammer}, \citenamefont {Allen},
  \citenamefont {Karlsson},\ and\ \citenamefont {Jepsen}}]{Gunnarson2001}%
  \BibitemOpen
  \bibfield  {author} {\bibinfo {author} {\bibfnamefont {O.}~\bibnamefont
  {Gunnarsson}}, \bibinfo {author} {\bibfnamefont {K.}~\bibnamefont
  {Sch\"onhammer}}, \bibinfo {author} {\bibfnamefont {J.}~\bibnamefont
  {Allen}}, \bibinfo {author} {\bibfnamefont {K.}~\bibnamefont {Karlsson}},\
  and\ \bibinfo {author} {\bibfnamefont {O.}~\bibnamefont {Jepsen}},\
  }\bibfield  {title} {\bibinfo {title} {Information from photoemission
  spectral weights and shapes},\ }\href
  {https://doi.org/10.1016/S0368-2048(01)00241-9} {\bibfield  {journal}
  {\bibinfo  {journal} {J. Electron Spectrosc. Relat. Phenom.}\ }\textbf
  {\bibinfo {volume} {117--118}},\ \bibinfo {pages} {1} (\bibinfo {year}
  {2001})}\BibitemShut {NoStop}%
\bibitem [{\citenamefont {Beyermann}\ \emph {et~al.}(1991)\citenamefont
  {Beyermann}, \citenamefont {Hundley}, \citenamefont {Canfield}, \citenamefont
  {Thompson}, \citenamefont {Latroche}, \citenamefont {Godart}, \citenamefont
  {Selsane}, \citenamefont {Fisk},\ and\ \citenamefont
  {Smith}}]{Beyermann1991}%
  \BibitemOpen
  \bibfield  {author} {\bibinfo {author} {\bibfnamefont {W.~P.}\ \bibnamefont
  {Beyermann}}, \bibinfo {author} {\bibfnamefont {M.~F.}\ \bibnamefont
  {Hundley}}, \bibinfo {author} {\bibfnamefont {P.~C.}\ \bibnamefont
  {Canfield}}, \bibinfo {author} {\bibfnamefont {J.~D.}\ \bibnamefont
  {Thompson}}, \bibinfo {author} {\bibfnamefont {M.}~\bibnamefont {Latroche}},
  \bibinfo {author} {\bibfnamefont {C.}~\bibnamefont {Godart}}, \bibinfo
  {author} {\bibfnamefont {M.}~\bibnamefont {Selsane}}, \bibinfo {author}
  {\bibfnamefont {Z.}~\bibnamefont {Fisk}},\ and\ \bibinfo {author}
  {\bibfnamefont {J.~L.}\ \bibnamefont {Smith}},\ }\bibfield  {title} {\bibinfo
  {title} {Competing interactions in the heavy-electron antiferromagnets
  {C}e${M}_2${Sn}$_2$ ({M=Ni, Ir, Cu, Rh, Pd, and Pt})},\ }\href
  {https://doi.org/10.1103/PhysRevB.43.13130} {\bibfield  {journal} {\bibinfo
  {journal} {Phys. Rev. B}\ }\textbf {\bibinfo {volume} {43}},\ \bibinfo
  {pages} {13130} (\bibinfo {year} {1991})}\BibitemShut {NoStop}%
\bibitem [{\citenamefont {Shimura}\ \emph {et~al.}(2021)\citenamefont
  {Shimura}, \citenamefont {W\"orl}, \citenamefont {Sundermann}, \citenamefont
  {Tsuda}, \citenamefont {Adroja}, \citenamefont {Bhattacharyya}, \citenamefont
  {Strydom}, \citenamefont {Hillier}, \citenamefont {Pratt}, \citenamefont
  {Gloskovskii}, \citenamefont {Severing}, \citenamefont {Onimaru},
  \citenamefont {Gegenwart},\ and\ \citenamefont {Takabatake}}]{Shimura2021}%
  \BibitemOpen
  \bibfield  {author} {\bibinfo {author} {\bibfnamefont {Y.}~\bibnamefont
  {Shimura}}, \bibinfo {author} {\bibfnamefont {A.}~\bibnamefont {W\"orl}},
  \bibinfo {author} {\bibfnamefont {M.}~\bibnamefont {Sundermann}}, \bibinfo
  {author} {\bibfnamefont {S.}~\bibnamefont {Tsuda}}, \bibinfo {author}
  {\bibfnamefont {D.~T.}\ \bibnamefont {Adroja}}, \bibinfo {author}
  {\bibfnamefont {A.}~\bibnamefont {Bhattacharyya}}, \bibinfo {author}
  {\bibfnamefont {A.~M.}\ \bibnamefont {Strydom}}, \bibinfo {author}
  {\bibfnamefont {A.~D.}\ \bibnamefont {Hillier}}, \bibinfo {author}
  {\bibfnamefont {F.~L.}\ \bibnamefont {Pratt}}, \bibinfo {author}
  {\bibfnamefont {A.}~\bibnamefont {Gloskovskii}}, \bibinfo {author}
  {\bibfnamefont {A.}~\bibnamefont {Severing}}, \bibinfo {author}
  {\bibfnamefont {T.}~\bibnamefont {Onimaru}}, \bibinfo {author} {\bibfnamefont
  {P.}~\bibnamefont {Gegenwart}},\ and\ \bibinfo {author} {\bibfnamefont
  {T.}~\bibnamefont {Takabatake}},\ }\bibfield  {title} {\bibinfo {title}
  {Antiferromagnetic correlations in strongly valence fluctuating {CeIrSn}},\
  }\href {https://doi.org/10.1103/PhysRevLett.126.217202} {\bibfield  {journal}
  {\bibinfo  {journal} {Phys. Rev. Lett.}\ }\textbf {\bibinfo {volume} {126}},\
  \bibinfo {pages} {217202} (\bibinfo {year} {2021})}\BibitemShut {NoStop}%
\bibitem [{\citenamefont {Allen}\ \emph {et~al.}(1986)\citenamefont {Allen},
  \citenamefont {Oh}, \citenamefont {Gunnarsson}, \citenamefont
  {Sch\"onhammer}, \citenamefont {Maple}, \citenamefont {Torikachvili},\ and\
  \citenamefont {Lindau}}]{Allen1986}%
  \BibitemOpen
  \bibfield  {author} {\bibinfo {author} {\bibfnamefont {J.}~\bibnamefont
  {Allen}}, \bibinfo {author} {\bibfnamefont {S.}~\bibnamefont {Oh}}, \bibinfo
  {author} {\bibfnamefont {O.}~\bibnamefont {Gunnarsson}}, \bibinfo {author}
  {\bibfnamefont {K.}~\bibnamefont {Sch\"onhammer}}, \bibinfo {author}
  {\bibfnamefont {M.}~\bibnamefont {Maple}}, \bibinfo {author} {\bibfnamefont
  {M.}~\bibnamefont {Torikachvili}},\ and\ \bibinfo {author} {\bibfnamefont
  {I.}~\bibnamefont {Lindau}},\ }\bibfield  {title} {\bibinfo {title}
  {Electronic structure of cerium and light rare-earth intermetallics},\ }\href
  {https://doi.org/10.1080/00018738600101901} {\bibfield  {journal} {\bibinfo
  {journal} {Adv. Phys.}\ }\textbf {\bibinfo {volume} {35}},\ \bibinfo {pages}
  {275} (\bibinfo {year} {1986})}\BibitemShut {NoStop}%
\bibitem [{\citenamefont {H{\"u}fner}(1992)}]{Huefner1992}%
  \BibitemOpen
  \bibfield  {author} {\bibinfo {author} {\bibfnamefont {S.}~\bibnamefont
  {H{\"u}fner}},\ }\bibfield  {title} {\bibinfo {title} {Valence band
  photoemission on metallic {C}e systems},\ }\href
  {https://doi.org/10.1007/BF01313831} {\bibfield  {journal} {\bibinfo
  {journal} {Z. Phys. B Cond. Mat.}\ }\textbf {\bibinfo {volume} {86}},\
  \bibinfo {pages} {241} (\bibinfo {year} {1992})}\BibitemShut {NoStop}%
\bibitem [{\citenamefont {Tjeng}\ \emph {et~al.}(1993)\citenamefont {Tjeng},
  \citenamefont {Oh}, \citenamefont {Cho}, \citenamefont {Lin}, \citenamefont
  {Chen}, \citenamefont {Gweon}, \citenamefont {Park}, \citenamefont {Allen},
  \citenamefont {Suzuki}, \citenamefont {Makivi\ifmmode~\acute{c}\else
  \'{c}\fi{}},\ and\ \citenamefont {Cox}}]{Tjeng1993}%
  \BibitemOpen
  \bibfield  {author} {\bibinfo {author} {\bibfnamefont {L.~H.}\ \bibnamefont
  {Tjeng}}, \bibinfo {author} {\bibfnamefont {S.-J.}\ \bibnamefont {Oh}},
  \bibinfo {author} {\bibfnamefont {E.-J.}\ \bibnamefont {Cho}}, \bibinfo
  {author} {\bibfnamefont {H.-J.}\ \bibnamefont {Lin}}, \bibinfo {author}
  {\bibfnamefont {C.~T.}\ \bibnamefont {Chen}}, \bibinfo {author}
  {\bibfnamefont {G.-H.}\ \bibnamefont {Gweon}}, \bibinfo {author}
  {\bibfnamefont {J.-H.}\ \bibnamefont {Park}}, \bibinfo {author}
  {\bibfnamefont {J.~W.}\ \bibnamefont {Allen}}, \bibinfo {author}
  {\bibfnamefont {T.}~\bibnamefont {Suzuki}}, \bibinfo {author} {\bibfnamefont
  {M.~S.}\ \bibnamefont {Makivi\ifmmode~\acute{c}\else \'{c}\fi{}}},\ and\
  \bibinfo {author} {\bibfnamefont {D.~L.}\ \bibnamefont {Cox}},\ }\bibfield
  {title} {\bibinfo {title} {Temperature dependence of the {K}ondo resonance in
  {YbAl}$_3$},\ }\href {https://doi.org/10.1103/PhysRevLett.71.1419} {\bibfield
   {journal} {\bibinfo  {journal} {Phys. Rev. Lett.}\ }\textbf {\bibinfo
  {volume} {71}},\ \bibinfo {pages} {1419} (\bibinfo {year}
  {1993})}\BibitemShut {NoStop}%
\bibitem [{\citenamefont {Braicovich}\ \emph {et~al.}(1997)\citenamefont
  {Braicovich}, \citenamefont {Brookes}, \citenamefont {Dallera}, \citenamefont
  {Salvietti},\ and\ \citenamefont {Olcese}}]{Braicovich1997}%
  \BibitemOpen
  \bibfield  {author} {\bibinfo {author} {\bibfnamefont {L.}~\bibnamefont
  {Braicovich}}, \bibinfo {author} {\bibfnamefont {N.~B.}\ \bibnamefont
  {Brookes}}, \bibinfo {author} {\bibfnamefont {C.}~\bibnamefont {Dallera}},
  \bibinfo {author} {\bibfnamefont {M.}~\bibnamefont {Salvietti}},\ and\
  \bibinfo {author} {\bibfnamefont {G.~L.}\ \bibnamefont {Olcese}},\ }\bibfield
   {title} {\bibinfo {title} {High-energy {Ce}-3$d$ photoemission: Bulk
  properties of {Ce$M_2$} (${M}$={Fe,Co,Ni}) and {Ce$_7$Ni}$_3$},\ }\href
  {https://doi.org/10.1103/PhysRevB.56.15047} {\bibfield  {journal} {\bibinfo
  {journal} {Phys. Rev. B}\ }\textbf {\bibinfo {volume} {56}},\ \bibinfo
  {pages} {15047} (\bibinfo {year} {1997})}\BibitemShut {NoStop}%
\bibitem [{\citenamefont {Dallera}\ \emph {et~al.}(2005)\citenamefont
  {Dallera}, \citenamefont {Braicovich}, \citenamefont {Duo}, \citenamefont
  {Palenzona}, \citenamefont {Panaccione}, \citenamefont {Paolicelli},
  \citenamefont {Cowie},\ and\ \citenamefont {Zegenhagen}}]{Dallera2005}%
  \BibitemOpen
  \bibfield  {author} {\bibinfo {author} {\bibfnamefont {C.}~\bibnamefont
  {Dallera}}, \bibinfo {author} {\bibfnamefont {L.}~\bibnamefont {Braicovich}},
  \bibinfo {author} {\bibfnamefont {L.}~\bibnamefont {Duo}}, \bibinfo {author}
  {\bibfnamefont {A.}~\bibnamefont {Palenzona}}, \bibinfo {author}
  {\bibfnamefont {G.}~\bibnamefont {Panaccione}}, \bibinfo {author}
  {\bibfnamefont {G.}~\bibnamefont {Paolicelli}}, \bibinfo {author}
  {\bibfnamefont {B.~C.~C.}\ \bibnamefont {Cowie}},\ and\ \bibinfo {author}
  {\bibfnamefont {J.}~\bibnamefont {Zegenhagen}},\ }\bibfield  {title}
  {\bibinfo {title} {Hard x-ray photoelectron spectroscopy: sensitivity to
  depth, chemistry and orbital character},\ }\href
  {https://doi.org/10.1016/j.nima.2005.05.017} {\bibfield  {journal} {\bibinfo
  {journal} {Nuc. Instr. and Meth. in Phys. Res. A}\ }\textbf {\bibinfo
  {volume} {547}},\ \bibinfo {pages} {155147} (\bibinfo {year}
  {2005})}\BibitemShut {NoStop}%
\bibitem [{\citenamefont {Yamasaki}\ \emph {et~al.}(2007)\citenamefont
  {Yamasaki}, \citenamefont {Imada}, \citenamefont {Higashimichi},
  \citenamefont {Fujiwara}, \citenamefont {Saita}, \citenamefont {Miyamachi},
  \citenamefont {Sekiyama}, \citenamefont {Sugawara}, \citenamefont {Kikuchi},
  \citenamefont {Sato}, \citenamefont {Higashiya}, \citenamefont {Yabashi},
  \citenamefont {Tamasaku}, \citenamefont {Miwa}, \citenamefont {Ishikawa},\
  and\ \citenamefont {Suga}}]{Suga_Sm_2007}%
  \BibitemOpen
  \bibfield  {author} {\bibinfo {author} {\bibfnamefont {A.}~\bibnamefont
  {Yamasaki}}, \bibinfo {author} {\bibfnamefont {S.}~\bibnamefont {Imada}},
  \bibinfo {author} {\bibfnamefont {H.}~\bibnamefont {Higashimichi}}, \bibinfo
  {author} {\bibfnamefont {H.}~\bibnamefont {Fujiwara}}, \bibinfo {author}
  {\bibfnamefont {T.}~\bibnamefont {Saita}}, \bibinfo {author} {\bibfnamefont
  {T.}~\bibnamefont {Miyamachi}}, \bibinfo {author} {\bibfnamefont
  {A.}~\bibnamefont {Sekiyama}}, \bibinfo {author} {\bibfnamefont
  {H.}~\bibnamefont {Sugawara}}, \bibinfo {author} {\bibfnamefont
  {D.}~\bibnamefont {Kikuchi}}, \bibinfo {author} {\bibfnamefont
  {H.}~\bibnamefont {Sato}}, \bibinfo {author} {\bibfnamefont {A.}~\bibnamefont
  {Higashiya}}, \bibinfo {author} {\bibfnamefont {M.}~\bibnamefont {Yabashi}},
  \bibinfo {author} {\bibfnamefont {K.}~\bibnamefont {Tamasaku}}, \bibinfo
  {author} {\bibfnamefont {D.}~\bibnamefont {Miwa}}, \bibinfo {author}
  {\bibfnamefont {T.}~\bibnamefont {Ishikawa}},\ and\ \bibinfo {author}
  {\bibfnamefont {S.}~\bibnamefont {Suga}},\ }\bibfield  {title} {\bibinfo
  {title} {Coexistence of strongly mixed-valence and heavy-fermion character in
  {SmOs}$_4${Sb}$_{12}$ studied by soft- and hard-x-ray spectroscopy},\ }\href
  {https://doi.org/10.1103/PhysRevLett.98.156402} {\bibfield  {journal}
  {\bibinfo  {journal} {Phys. Rev. Lett.}\ }\textbf {\bibinfo {volume} {98}},\
  \bibinfo {pages} {156402} (\bibinfo {year} {2007})}\BibitemShut {NoStop}%
\bibitem [{\citenamefont {Tanuma}\ \emph {et~al.}(2011)\citenamefont {Tanuma},
  \citenamefont {Powell},\ and\ \citenamefont {Penn}}]{TanumaIMFP_2011}%
  \BibitemOpen
  \bibfield  {author} {\bibinfo {author} {\bibfnamefont {S.}~\bibnamefont
  {Tanuma}}, \bibinfo {author} {\bibfnamefont {C.~J.}\ \bibnamefont {Powell}},\
  and\ \bibinfo {author} {\bibfnamefont {D.~R.}\ \bibnamefont {Penn}},\
  }\bibfield  {title} {\bibinfo {title} {Calculations of electron inelastic
  mean free paths. {IX} {D}ata for 41 elemental solids over the 50 e{V} to 30
  ke{V} range},\ }\href {https://doi.org/10.1002/sia.3522} {\bibfield
  {journal} {\bibinfo  {journal} {Surface and Interface Analysis}\ }\textbf
  {\bibinfo {volume} {43}},\ \bibinfo {pages} {689} (\bibinfo {year}
  {2011})}\BibitemShut {NoStop}%
\bibitem [{\citenamefont {Laubschat}\ \emph {et~al.}(1990)\citenamefont
  {Laubschat}, \citenamefont {Weschke}, \citenamefont {Holtz}, \citenamefont
  {Domke}, \citenamefont {Strebel},\ and\ \citenamefont
  {Kaindl}}]{Laubschat1990}%
  \BibitemOpen
  \bibfield  {author} {\bibinfo {author} {\bibfnamefont {C.}~\bibnamefont
  {Laubschat}}, \bibinfo {author} {\bibfnamefont {E.}~\bibnamefont {Weschke}},
  \bibinfo {author} {\bibfnamefont {C.}~\bibnamefont {Holtz}}, \bibinfo
  {author} {\bibfnamefont {M.}~\bibnamefont {Domke}}, \bibinfo {author}
  {\bibfnamefont {O.}~\bibnamefont {Strebel}},\ and\ \bibinfo {author}
  {\bibfnamefont {G.}~\bibnamefont {Kaindl}},\ }\bibfield  {title} {\bibinfo
  {title} {Surface-electronic structure of $\alpha${}-like {C}e compounds},\
  }\href {https://doi.org/10.1103/PhysRevLett.65.1639} {\bibfield  {journal}
  {\bibinfo  {journal} {Phys. Rev. Lett.}\ }\textbf {\bibinfo {volume} {65}},\
  \bibinfo {pages} {1639} (\bibinfo {year} {1990})}\BibitemShut {NoStop}%
\bibitem [{\citenamefont {Sekiyama}\ \emph {et~al.}(2000)\citenamefont
  {Sekiyama}, \citenamefont {Iwasaki}, \citenamefont {Matsuda}, \citenamefont
  {Saitoh}, \citenamefont {Onuki},\ and\ \citenamefont {Suga}}]{Suga_Ce_2000}%
  \BibitemOpen
  \bibfield  {author} {\bibinfo {author} {\bibfnamefont {A.}~\bibnamefont
  {Sekiyama}}, \bibinfo {author} {\bibfnamefont {T.}~\bibnamefont {Iwasaki}},
  \bibinfo {author} {\bibfnamefont {K.}~\bibnamefont {Matsuda}}, \bibinfo
  {author} {\bibfnamefont {Y.}~\bibnamefont {Saitoh}}, \bibinfo {author}
  {\bibfnamefont {Y.}~\bibnamefont {Onuki}},\ and\ \bibinfo {author}
  {\bibfnamefont {S.}~\bibnamefont {Suga}},\ }\bibfield  {title} {\bibinfo
  {title} {Probing bulk states of correlated electron systems by
  high-resolution resonance photoemission},\ }\href
  {https://doi.org/10.1038/35000140} {\bibfield  {journal} {\bibinfo  {journal}
  {Nature}\ }\textbf {\bibinfo {volume} {403}},\ \bibinfo {pages} {396}
  (\bibinfo {year} {2000})}\BibitemShut {NoStop}%
\bibitem [{\citenamefont {Schlueter}\ \emph {et~al.}(2019)\citenamefont
  {Schlueter}, \citenamefont {Gloskovskii}, \citenamefont {Ederer},
  \citenamefont {Schostak}, \citenamefont {Piec}, \citenamefont {Sarkar},
  \citenamefont {Matveyev}, \citenamefont {L\"omker}, \citenamefont {Sing},
  \citenamefont {Claessen}, \citenamefont {Wiemann}, \citenamefont {Schneider},
  \citenamefont {Medjanik}, \citenamefont {Sch\"onhense}, \citenamefont
  {Amann}, \citenamefont {Nilsson},\ and\ \citenamefont
  {Drube}}]{Schlueter2019}%
  \BibitemOpen
  \bibfield  {author} {\bibinfo {author} {\bibfnamefont {C.}~\bibnamefont
  {Schlueter}}, \bibinfo {author} {\bibfnamefont {A.}~\bibnamefont
  {Gloskovskii}}, \bibinfo {author} {\bibfnamefont {K.}~\bibnamefont {Ederer}},
  \bibinfo {author} {\bibfnamefont {I.}~\bibnamefont {Schostak}}, \bibinfo
  {author} {\bibfnamefont {S.}~\bibnamefont {Piec}}, \bibinfo {author}
  {\bibfnamefont {I.}~\bibnamefont {Sarkar}}, \bibinfo {author} {\bibfnamefont
  {Y.}~\bibnamefont {Matveyev}}, \bibinfo {author} {\bibfnamefont
  {P.}~\bibnamefont {L\"omker}}, \bibinfo {author} {\bibfnamefont
  {M.}~\bibnamefont {Sing}}, \bibinfo {author} {\bibfnamefont {R.}~\bibnamefont
  {Claessen}}, \bibinfo {author} {\bibfnamefont {C.}~\bibnamefont {Wiemann}},
  \bibinfo {author} {\bibfnamefont {C.~M.}\ \bibnamefont {Schneider}}, \bibinfo
  {author} {\bibfnamefont {K.}~\bibnamefont {Medjanik}}, \bibinfo {author}
  {\bibfnamefont {G.}~\bibnamefont {Sch\"onhense}}, \bibinfo {author}
  {\bibfnamefont {P.}~\bibnamefont {Amann}}, \bibinfo {author} {\bibfnamefont
  {A.}~\bibnamefont {Nilsson}},\ and\ \bibinfo {author} {\bibfnamefont
  {W.}~\bibnamefont {Drube}},\ }\bibfield  {title} {\bibinfo {title} {The new
  dedicated {HAXPES} beamline {P22} at {PETRA\,III}},\ }\href
  {https://doi.org/10.1063/1.5084611} {\bibfield  {journal} {\bibinfo
  {journal} {AIP Conf. Proc.}\ }\textbf {\bibinfo {volume} {2054}},\ \bibinfo
  {pages} {040010} (\bibinfo {year} {2019})}\BibitemShut {NoStop}%
\bibitem [{\citenamefont {Amorese}\ \emph {et~al.}(2020)\citenamefont
  {Amorese}, \citenamefont {Sundermann}, \citenamefont {Leedahl}, \citenamefont
  {Marino}, \citenamefont {Takegami}, \citenamefont {Gretarsson}, \citenamefont
  {Gloskovskii}, \citenamefont {Schlueter}, \citenamefont {Haverkort},
  \citenamefont {Huang}, \citenamefont {Szlawska}, \citenamefont {Kaczorowski},
  \citenamefont {Ran}, \citenamefont {Maple}, \citenamefont {Bauer},
  \citenamefont {Leithe-Jasper}, \citenamefont {Hansmann}, \citenamefont
  {Thalmeier}, \citenamefont {Tjeng},\ and\ \citenamefont
  {Severing}}]{Amorese2020}%
  \BibitemOpen
  \bibfield  {author} {\bibinfo {author} {\bibfnamefont {A.}~\bibnamefont
  {Amorese}}, \bibinfo {author} {\bibfnamefont {M.}~\bibnamefont {Sundermann}},
  \bibinfo {author} {\bibfnamefont {B.}~\bibnamefont {Leedahl}}, \bibinfo
  {author} {\bibfnamefont {A.}~\bibnamefont {Marino}}, \bibinfo {author}
  {\bibfnamefont {D.}~\bibnamefont {Takegami}}, \bibinfo {author}
  {\bibfnamefont {H.}~\bibnamefont {Gretarsson}}, \bibinfo {author}
  {\bibfnamefont {A.}~\bibnamefont {Gloskovskii}}, \bibinfo {author}
  {\bibfnamefont {C.}~\bibnamefont {Schlueter}}, \bibinfo {author}
  {\bibfnamefont {M.~W.}\ \bibnamefont {Haverkort}}, \bibinfo {author}
  {\bibfnamefont {Y.}~\bibnamefont {Huang}}, \bibinfo {author} {\bibfnamefont
  {M.}~\bibnamefont {Szlawska}}, \bibinfo {author} {\bibfnamefont
  {D.}~\bibnamefont {Kaczorowski}}, \bibinfo {author} {\bibfnamefont
  {S.}~\bibnamefont {Ran}}, \bibinfo {author} {\bibfnamefont {M.~B.}\
  \bibnamefont {Maple}}, \bibinfo {author} {\bibfnamefont {E.~D.}\ \bibnamefont
  {Bauer}}, \bibinfo {author} {\bibfnamefont {A.}~\bibnamefont
  {Leithe-Jasper}}, \bibinfo {author} {\bibfnamefont {P.}~\bibnamefont
  {Hansmann}}, \bibinfo {author} {\bibfnamefont {P.}~\bibnamefont {Thalmeier}},
  \bibinfo {author} {\bibfnamefont {L.~H.}\ \bibnamefont {Tjeng}},\ and\
  \bibinfo {author} {\bibfnamefont {A.}~\bibnamefont {Severing}},\ }\bibfield
  {title} {\bibinfo {title} {From antiferromagnetic and hidden order to {P}auli
  paramagnetism in {U$M_2$S}i$_2$ compounds with 5$f$ electron duality},\
  }\href {https://doi.org/10.1073/pnas.2005701117} {\bibfield  {journal}
  {\bibinfo  {journal} {Proc. Nat. Acad. Sci.}\ }\textbf {\bibinfo {volume}
  {117}},\ \bibinfo {pages} {30220} (\bibinfo {year} {2020})}\BibitemShut
  {NoStop}%
\bibitem [{\citenamefont {Tanaka}\ and\ \citenamefont {Jo}(1994)}]{Tanaka1994}%
  \BibitemOpen
  \bibfield  {author} {\bibinfo {author} {\bibfnamefont {A.}~\bibnamefont
  {Tanaka}}\ and\ \bibinfo {author} {\bibfnamefont {T.}~\bibnamefont {Jo}},\
  }\href@noop {} {\bibfield  {journal} {\bibinfo  {journal} {J. Phys. Soc.
  Japan}\ }\textbf {\bibinfo {volume} {63}},\ \bibinfo {pages} {2788} (\bibinfo
  {year} {1994})}\BibitemShut {NoStop}%
\bibitem [{\citenamefont {Imer}\ and\ \citenamefont
  {Wuilloud}(1987)}]{Imer1987}%
  \BibitemOpen
  \bibfield  {author} {\bibinfo {author} {\bibfnamefont {J.~M.}\ \bibnamefont
  {Imer}}\ and\ \bibinfo {author} {\bibfnamefont {E.}~\bibnamefont
  {Wuilloud}},\ }\bibfield  {title} {\bibinfo {title} {A simple model
  calculation for {XPS, BIS and EELS} 4f-excitations in {C}e and {L}a
  compounds},\ }\href {https://doi.org/10.1007/BF01311650} {\bibfield
  {journal} {\bibinfo  {journal} {Z. Physik B Cond. Matt.}\ }\textbf {\bibinfo
  {volume} {66}},\ \bibinfo {pages} {153} (\bibinfo {year} {1987})}\BibitemShut
  {NoStop}%
\bibitem [{\citenamefont {Strigari}\ \emph {et~al.}(2015)\citenamefont
  {Strigari}, \citenamefont {Sundermann}, \citenamefont {Muro}, \citenamefont
  {Yutani}, \citenamefont {Takabatake}, \citenamefont {Tsuei}, \citenamefont
  {Liao}, \citenamefont {Tanaka}, \citenamefont {Thalmeier}, \citenamefont
  {Haverkort}, \citenamefont {Tjeng},\ and\ \citenamefont
  {Severing}}]{Strigari2015}%
  \BibitemOpen
  \bibfield  {author} {\bibinfo {author} {\bibfnamefont {F.}~\bibnamefont
  {Strigari}}, \bibinfo {author} {\bibfnamefont {M.}~\bibnamefont
  {Sundermann}}, \bibinfo {author} {\bibfnamefont {Y.}~\bibnamefont {Muro}},
  \bibinfo {author} {\bibfnamefont {K.}~\bibnamefont {Yutani}}, \bibinfo
  {author} {\bibfnamefont {T.}~\bibnamefont {Takabatake}}, \bibinfo {author}
  {\bibfnamefont {K.-D.}\ \bibnamefont {Tsuei}}, \bibinfo {author}
  {\bibfnamefont {Y.}~\bibnamefont {Liao}}, \bibinfo {author} {\bibfnamefont
  {A.}~\bibnamefont {Tanaka}}, \bibinfo {author} {\bibfnamefont
  {P.}~\bibnamefont {Thalmeier}}, \bibinfo {author} {\bibfnamefont
  {M.}~\bibnamefont {Haverkort}}, \bibinfo {author} {\bibfnamefont
  {L.}~\bibnamefont {Tjeng}},\ and\ \bibinfo {author} {\bibfnamefont
  {A.}~\bibnamefont {Severing}},\ }\bibfield  {title} {\bibinfo {title}
  {Quantitative study of valence and configuration interaction parameters of
  the {K}ondo semiconductors {Ce$M_2$Al}$_{10}$ ({$M$=Ru, Os and Fe}) by means
  of bulk-sensitive hard x-ray photoelectron spectroscopy},\ }\href
  {https://doi.org/https://doi.org/10.1016/j.elspec.2015.01.004} {\bibfield
  {journal} {\bibinfo  {journal} {J. Elec. Spect. and Rel. Phen.}\ }\textbf
  {\bibinfo {volume} {199}},\ \bibinfo {pages} {56} (\bibinfo {year}
  {2015})}\BibitemShut {NoStop}%
\bibitem [{\citenamefont {Sundermann}\ \emph {et~al.}(2016)\citenamefont
  {Sundermann}, \citenamefont {Strigari}, \citenamefont {Willers},
  \citenamefont {Weinen}, \citenamefont {Liao}, \citenamefont {Tsuei},
  \citenamefont {Hiraoka}, \citenamefont {Ishii}, \citenamefont {Yamaoka},
  \citenamefont {Mizuki}, \citenamefont {Zekko}, \citenamefont {Bauer},
  \citenamefont {Sarrao}, \citenamefont {Thompson}, \citenamefont {Lejay},
  \citenamefont {Muro}, \citenamefont {Yutani}, \citenamefont {Takabatake},
  \citenamefont {Tanaka}, \citenamefont {Hollmann}, \citenamefont {Tjeng},\
  and\ \citenamefont {Severing}}]{Sundermann2016}%
  \BibitemOpen
  \bibfield  {author} {\bibinfo {author} {\bibfnamefont {M.}~\bibnamefont
  {Sundermann}}, \bibinfo {author} {\bibfnamefont {F.}~\bibnamefont
  {Strigari}}, \bibinfo {author} {\bibfnamefont {T.}~\bibnamefont {Willers}},
  \bibinfo {author} {\bibfnamefont {J.}~\bibnamefont {Weinen}}, \bibinfo
  {author} {\bibfnamefont {Y.}~\bibnamefont {Liao}}, \bibinfo {author}
  {\bibfnamefont {K.-D.}\ \bibnamefont {Tsuei}}, \bibinfo {author}
  {\bibfnamefont {N.}~\bibnamefont {Hiraoka}}, \bibinfo {author} {\bibfnamefont
  {H.}~\bibnamefont {Ishii}}, \bibinfo {author} {\bibfnamefont
  {H.}~\bibnamefont {Yamaoka}}, \bibinfo {author} {\bibfnamefont
  {J.}~\bibnamefont {Mizuki}}, \bibinfo {author} {\bibfnamefont
  {Y.}~\bibnamefont {Zekko}}, \bibinfo {author} {\bibfnamefont
  {E.}~\bibnamefont {Bauer}}, \bibinfo {author} {\bibfnamefont
  {J.}~\bibnamefont {Sarrao}}, \bibinfo {author} {\bibfnamefont
  {J.}~\bibnamefont {Thompson}}, \bibinfo {author} {\bibfnamefont
  {P.}~\bibnamefont {Lejay}}, \bibinfo {author} {\bibfnamefont
  {Y.}~\bibnamefont {Muro}}, \bibinfo {author} {\bibfnamefont {K.}~\bibnamefont
  {Yutani}}, \bibinfo {author} {\bibfnamefont {T.}~\bibnamefont {Takabatake}},
  \bibinfo {author} {\bibfnamefont {A.}~\bibnamefont {Tanaka}}, \bibinfo
  {author} {\bibfnamefont {N.}~\bibnamefont {Hollmann}}, \bibinfo {author}
  {\bibfnamefont {L.}~\bibnamefont {Tjeng}},\ and\ \bibinfo {author}
  {\bibfnamefont {A.}~\bibnamefont {Severing}},\ }\bibfield  {title} {\bibinfo
  {title} {Quantitative study of the $f$ occupation in {Ce$M$In$_5$} and other
  cerium compounds with hard x-rays},\ }\href
  {https://doi.org/https://doi.org/10.1016/j.elspec.2016.02.002} {\bibfield
  {journal} {\bibinfo  {journal} {J. Elec. Spect. and Rel. Phen.}\ }\textbf
  {\bibinfo {volume} {209}},\ \bibinfo {pages} {1} (\bibinfo {year}
  {2016})}\BibitemShut {NoStop}%
\bibitem [{\citenamefont {Koepernik}\ and\ \citenamefont
  {Eschrig}(1999)}]{Koepernik1999}%
  \BibitemOpen
  \bibfield  {author} {\bibinfo {author} {\bibfnamefont {K.}~\bibnamefont
  {Koepernik}}\ and\ \bibinfo {author} {\bibfnamefont {H.}~\bibnamefont
  {Eschrig}},\ }\bibfield  {title} {\bibinfo {title} {Full-potential
  nonorthogonal local-orbital minimum-basis band-structure scheme},\ }\href
  {https://doi.org/10.1103/PhysRevB.59.1743} {\bibfield  {journal} {\bibinfo
  {journal} {Phys. Rev. B}\ }\textbf {\bibinfo {volume} {59}},\ \bibinfo
  {pages} {1743} (\bibinfo {year} {1999})}\BibitemShut {NoStop}%
\bibitem [{\citenamefont {Trzhaskovskaya}\ \emph {et~al.}(2006)\citenamefont
  {Trzhaskovskaya}, \citenamefont {Nikulin}, \citenamefont {Nefedov},\ and\
  \citenamefont {Yarzhemsky}}]{TRZHASKOVSKAYA2006245}%
  \BibitemOpen
  \bibfield  {author} {\bibinfo {author} {\bibfnamefont {M.~B.}\ \bibnamefont
  {Trzhaskovskaya}}, \bibinfo {author} {\bibfnamefont {V.~K.}\ \bibnamefont
  {Nikulin}}, \bibinfo {author} {\bibfnamefont {V.~I.}\ \bibnamefont
  {Nefedov}},\ and\ \bibinfo {author} {\bibfnamefont {V.~G.}\ \bibnamefont
  {Yarzhemsky}},\ }\bibfield  {title} {\bibinfo {title} {Non-dipole second
  order parameters of the photoelectron angular distribution for elements
  {Z}=1–100 in the photoelectron energy range 1–10{keV}},\ }\href
  {https://doi.org/https://doi.org/10.1016/j.adt.2005.12.002} {\bibfield
  {journal} {\bibinfo  {journal} {Atomic Data and Nuclear Data Tables}\
  }\textbf {\bibinfo {volume} {92}},\ \bibinfo {pages} {245 } (\bibinfo {year}
  {2006})}\BibitemShut {NoStop}%
\bibitem [{\citenamefont {Takegami}\ \emph {et~al.}(2019)\citenamefont
  {Takegami}, \citenamefont {Nicola\"{\i}}, \citenamefont {Koethe},
  \citenamefont {Kasinathan}, \citenamefont {Kuo}, \citenamefont {Liao},
  \citenamefont {Tsuei}, \citenamefont {Panaccione}, \citenamefont {Offi},
  \citenamefont {Monaco}, \citenamefont {Brookes}, \citenamefont {Min\'ar},\
  and\ \citenamefont {Tjeng}}]{Takegami2019}%
  \BibitemOpen
  \bibfield  {author} {\bibinfo {author} {\bibfnamefont {D.}~\bibnamefont
  {Takegami}}, \bibinfo {author} {\bibfnamefont {L.}~\bibnamefont
  {Nicola\"{\i}}}, \bibinfo {author} {\bibfnamefont {T.~C.}\ \bibnamefont
  {Koethe}}, \bibinfo {author} {\bibfnamefont {D.}~\bibnamefont {Kasinathan}},
  \bibinfo {author} {\bibfnamefont {C.~Y.}\ \bibnamefont {Kuo}}, \bibinfo
  {author} {\bibfnamefont {Y.~F.}\ \bibnamefont {Liao}}, \bibinfo {author}
  {\bibfnamefont {K.~D.}\ \bibnamefont {Tsuei}}, \bibinfo {author}
  {\bibfnamefont {G.}~\bibnamefont {Panaccione}}, \bibinfo {author}
  {\bibfnamefont {F.}~\bibnamefont {Offi}}, \bibinfo {author} {\bibfnamefont
  {G.}~\bibnamefont {Monaco}}, \bibinfo {author} {\bibfnamefont {N.~B.}\
  \bibnamefont {Brookes}}, \bibinfo {author} {\bibfnamefont {J.}~\bibnamefont
  {Min\'ar}},\ and\ \bibinfo {author} {\bibfnamefont {L.~H.}\ \bibnamefont
  {Tjeng}},\ }\bibfield  {title} {\bibinfo {title} {Valence band hard x-ray
  photoelectron spectroscopy on $3d$ transition-metal oxides containing
  rare-earth elements},\ }\href {https://doi.org/10.1103/PhysRevB.99.165101}
  {\bibfield  {journal} {\bibinfo  {journal} {Phys. Rev. B}\ }\textbf {\bibinfo
  {volume} {99}},\ \bibinfo {pages} {165101} (\bibinfo {year}
  {2019})}\BibitemShut {NoStop}%
\bibitem [{\citenamefont {Sundermann}\ \emph {et~al.}(2017)\citenamefont
  {Sundermann}, \citenamefont {Chen}, \citenamefont {Utsumi}, \citenamefont
  {Wu}, \citenamefont {Tsuei}, \citenamefont {Haenel}, \citenamefont
  {Prokofiev}, \citenamefont {Paschen}, \citenamefont {Tanaka}, \citenamefont
  {Tjeng},\ and\ \citenamefont {Severing}}]{Sundermann2017}%
  \BibitemOpen
  \bibfield  {author} {\bibinfo {author} {\bibfnamefont {M.}~\bibnamefont
  {Sundermann}}, \bibinfo {author} {\bibfnamefont {K.}~\bibnamefont {Chen}},
  \bibinfo {author} {\bibfnamefont {Y.}~\bibnamefont {Utsumi}}, \bibinfo
  {author} {\bibfnamefont {Y.-H.}\ \bibnamefont {Wu}}, \bibinfo {author}
  {\bibfnamefont {K.-D.}\ \bibnamefont {Tsuei}}, \bibinfo {author}
  {\bibfnamefont {J.}~\bibnamefont {Haenel}}, \bibinfo {author} {\bibfnamefont
  {A.}~\bibnamefont {Prokofiev}}, \bibinfo {author} {\bibfnamefont
  {S.}~\bibnamefont {Paschen}}, \bibinfo {author} {\bibfnamefont
  {A.}~\bibnamefont {Tanaka}}, \bibinfo {author} {\bibfnamefont {L.~H.}\
  \bibnamefont {Tjeng}},\ and\ \bibinfo {author} {\bibfnamefont
  {A.}~\bibnamefont {Severing}},\ }\bibfield  {title} {\bibinfo {title} {Ce
  3$p$ hard x-ray photoelectron spectroscopy study of the topological {K}ondo
  insulator {CeRu}$_4${S}n$_6$},\ }\href
  {https://doi.org/10.1088/1742-6596/807/2/022001} {\bibfield  {journal}
  {\bibinfo  {journal} {J. Phys.: Conf. Series}\ }\textbf {\bibinfo {volume}
  {807}},\ \bibinfo {pages} {022001} (\bibinfo {year} {2017})}\BibitemShut
  {NoStop}%
\bibitem [{\citenamefont {Gamża}\ \emph {et~al.}(2009)\citenamefont {Gamża},
  \citenamefont {Slebarski},\ and\ \citenamefont {Rosner}}]{Gamza2009}%
  \BibitemOpen
  \bibfield  {author} {\bibinfo {author} {\bibfnamefont {M.}~\bibnamefont
  {Gamża}}, \bibinfo {author} {\bibfnamefont {A.}~\bibnamefont {Slebarski}},\
  and\ \bibinfo {author} {\bibfnamefont {H.}~\bibnamefont {Rosner}},\
  }\bibfield  {title} {\bibinfo {title} {Electronic structure of {CeRhX´} {(X
  = Sn, In)}},\ }\href {https://doi.org/10.1140/epjb/e2009-00047-1} {\bibfield
  {journal} {\bibinfo  {journal} {The European Physical Journal B}\ }\textbf
  {\bibinfo {volume} {67}},\ \bibinfo {pages} {483} (\bibinfo {year}
  {2009})}\BibitemShut {NoStop}%
\bibitem [{\citenamefont {Murani}\ \emph {et~al.}(1993)\citenamefont {Murani},
  \citenamefont {Bowden}, \citenamefont {Taylor}, \citenamefont {Osborn},\ and\
  \citenamefont {Marshall}}]{Murani1993}%
  \BibitemOpen
  \bibfield  {author} {\bibinfo {author} {\bibfnamefont {A.~P.}\ \bibnamefont
  {Murani}}, \bibinfo {author} {\bibfnamefont {Z.~A.}\ \bibnamefont {Bowden}},
  \bibinfo {author} {\bibfnamefont {A.~D.}\ \bibnamefont {Taylor}}, \bibinfo
  {author} {\bibfnamefont {R.}~\bibnamefont {Osborn}},\ and\ \bibinfo {author}
  {\bibfnamefont {W.~G.}\ \bibnamefont {Marshall}},\ }\bibfield  {title}
  {\bibinfo {title} {Evidence for localized 4$f$ states in $\alpha$-{C}e},\
  }\href {https://doi.org/10.1103/PhysRevB.48.13981} {\bibfield  {journal}
  {\bibinfo  {journal} {Phys. Rev. B}\ }\textbf {\bibinfo {volume} {48}},\
  \bibinfo {pages} {13981} (\bibinfo {year} {1993})}\BibitemShut {NoStop}%
\bibitem [{\citenamefont {Kittaka}\ \emph {et~al.}(2021)\citenamefont
  {Kittaka}, \citenamefont {Kono}, \citenamefont {Tsuda}, \citenamefont
  {Takabatake},\ and\ \citenamefont {Sakakibara}}]{Kittaka2021}%
  \BibitemOpen
  \bibfield  {author} {\bibinfo {author} {\bibfnamefont {S.}~\bibnamefont
  {Kittaka}}, \bibinfo {author} {\bibfnamefont {Y.}~\bibnamefont {Kono}},
  \bibinfo {author} {\bibfnamefont {S.}~\bibnamefont {Tsuda}}, \bibinfo
  {author} {\bibfnamefont {T.}~\bibnamefont {Takabatake}},\ and\ \bibinfo
  {author} {\bibfnamefont {T.}~\bibnamefont {Sakakibara}},\ }\bibfield  {title}
  {\bibinfo {title} {Field-angle-resolved landscape of non-fermi-liquid
  behavior in the quasi-kagome {K}ondo lattice {CeRhSn}},\ }\href
  {https://doi.org/10.7566/JPSJ.90.064703} {\bibfield  {journal} {\bibinfo
  {journal} {J. Phys. Soc. Jpn.}\ }\textbf {\bibinfo {volume} {90}},\ \bibinfo
  {pages} {064703} (\bibinfo {year} {2021})}\BibitemShut {NoStop}%
\bibitem [{\citenamefont {Kohgi}\ \emph {et~al.}(1993)\citenamefont {Kohgi},
  \citenamefont {Ohoyama}, \citenamefont {Osakabe}, \citenamefont {Kasaya},
  \citenamefont {Takabatake},\ and\ \citenamefont {Fujii}}]{Kohgi1993}%
  \BibitemOpen
  \bibfield  {author} {\bibinfo {author} {\bibfnamefont {M.}~\bibnamefont
  {Kohgi}}, \bibinfo {author} {\bibfnamefont {K.}~\bibnamefont {Ohoyama}},
  \bibinfo {author} {\bibfnamefont {T.}~\bibnamefont {Osakabe}}, \bibinfo
  {author} {\bibfnamefont {M.}~\bibnamefont {Kasaya}}, \bibinfo {author}
  {\bibfnamefont {T.}~\bibnamefont {Takabatake}},\ and\ \bibinfo {author}
  {\bibfnamefont {H.}~\bibnamefont {Fujii}},\ }\bibfield  {title} {\bibinfo
  {title} {Neutron scattering study of {CeNiSn}},\ }\href
  {https://doi.org/https://doi.org/10.1016/0921-4526(93)90588-W} {\bibfield
  {journal} {\bibinfo  {journal} {Physica B: Cond. Matt.}\ }\textbf {\bibinfo
  {volume} {186-188}},\ \bibinfo {pages} {409} (\bibinfo {year}
  {1993})}\BibitemShut {NoStop}%
\bibitem [{\citenamefont {Tsuda}\ \emph {et~al.}(2018)\citenamefont {Tsuda},
  \citenamefont {Yang}, \citenamefont {Shimura}, \citenamefont {Umeo},
  \citenamefont {Fukuoka}, \citenamefont {Yamane}, \citenamefont {Onimaru},
  \citenamefont {Takabatake}, \citenamefont {Kikugawa}, \citenamefont
  {Terashima}, \citenamefont {Hirose}, \citenamefont {Uji}, \citenamefont
  {Kittaka},\ and\ \citenamefont {Sakakibara}}]{Tsuda2018}%
  \BibitemOpen
  \bibfield  {author} {\bibinfo {author} {\bibfnamefont {S.}~\bibnamefont
  {Tsuda}}, \bibinfo {author} {\bibfnamefont {C.~L.}\ \bibnamefont {Yang}},
  \bibinfo {author} {\bibfnamefont {Y.}~\bibnamefont {Shimura}}, \bibinfo
  {author} {\bibfnamefont {K.}~\bibnamefont {Umeo}}, \bibinfo {author}
  {\bibfnamefont {H.}~\bibnamefont {Fukuoka}}, \bibinfo {author} {\bibfnamefont
  {Y.}~\bibnamefont {Yamane}}, \bibinfo {author} {\bibfnamefont
  {T.}~\bibnamefont {Onimaru}}, \bibinfo {author} {\bibfnamefont
  {T.}~\bibnamefont {Takabatake}}, \bibinfo {author} {\bibfnamefont
  {N.}~\bibnamefont {Kikugawa}}, \bibinfo {author} {\bibfnamefont
  {T.}~\bibnamefont {Terashima}}, \bibinfo {author} {\bibfnamefont {H.~T.}\
  \bibnamefont {Hirose}}, \bibinfo {author} {\bibfnamefont {S.}~\bibnamefont
  {Uji}}, \bibinfo {author} {\bibfnamefont {S.}~\bibnamefont {Kittaka}},\ and\
  \bibinfo {author} {\bibfnamefont {T.}~\bibnamefont {Sakakibara}},\ }\bibfield
   {title} {\bibinfo {title} {Metamagnetic crossover in the quasikagome ising
  {K}ondo-lattice compound {CeIrSn}},\ }\href
  {https://doi.org/10.1103/PhysRevB.98.155147} {\bibfield  {journal} {\bibinfo
  {journal} {Phys. Rev. B}\ }\textbf {\bibinfo {volume} {98}},\ \bibinfo
  {pages} {155147} (\bibinfo {year} {2018})}\BibitemShut {NoStop}%
\bibitem [{\citenamefont {Moulder}\ \emph {et~al.}(1995)\citenamefont
  {Moulder}, \citenamefont {Stickler}, \citenamefont {Sobol},\ and\
  \citenamefont {Bomben}}]{Handbook}%
  \BibitemOpen
  \bibfield  {author} {\bibinfo {author} {\bibfnamefont {J.~F.}\ \bibnamefont
  {Moulder}}, \bibinfo {author} {\bibfnamefont {W.~F.}\ \bibnamefont
  {Stickler}}, \bibinfo {author} {\bibfnamefont {P.~E.}\ \bibnamefont
  {Sobol}},\ and\ \bibinfo {author} {\bibfnamefont {K.~D.}\ \bibnamefont
  {Bomben}},\ }\href@noop {} {\emph {\bibinfo {title} {Handbook of {X}-ray
  {P}hotoelectron {S}pectroscopy}}},\ edited by\ \bibinfo {editor}
  {\bibfnamefont {J.}~\bibnamefont {Chastain}}\ and\ \bibinfo {editor}
  {\bibfnamefont {J.}~\bibnamefont {Roger C.~King}}\ (\bibinfo  {publisher}
  {Physical Electronics. Inc},\ \bibinfo {year} {1995})\BibitemShut {NoStop}%
\bibitem [{\citenamefont {Mahan}(1975)}]{Mahan1975}%
  \BibitemOpen
  \bibfield  {author} {\bibinfo {author} {\bibfnamefont {G.~D.}\ \bibnamefont
  {Mahan}},\ }\bibfield  {title} {\bibinfo {title} {Collective excitations in
  x-ray spectra of metals},\ }\href {https://doi.org/10.1103/PhysRevB.11.4814}
  {\bibfield  {journal} {\bibinfo  {journal} {Phys. Rev. B}\ }\textbf {\bibinfo
  {volume} {11}},\ \bibinfo {pages} {4814} (\bibinfo {year}
  {1975})}\BibitemShut {NoStop}%
\end{thebibliography}
\end{document}